\documentclass[lettersize,journal]{IEEEtran}
\IEEEoverridecommandlockouts
\usepackage{makecell}
\usepackage{array}
\usepackage{graphicx,amssymb,amsmath}
\usepackage{multicol}
\usepackage[noadjust]{cite}
\usepackage{setspace}
\usepackage{subfigure}
\usepackage{graphicx}
\usepackage{float}
\usepackage{url}
\usepackage{stfloats}
\usepackage{amsthm,pifont}
\usepackage{flushend}
\usepackage{cases,subeqnarray}
\usepackage{bm,multirow,bigstrut}
\usepackage{amsmath, amsthm, amssymb}
\usepackage{textcomp}
\usepackage{latexsym,bm}
\usepackage{booktabs}
\usepackage{xcolor}
\usepackage{mathtools}
\usepackage{dsfont}
\usepackage{extarrows}
\usepackage{epsfig}
\usepackage{epstopdf}
\usepackage[noend]{algpseudocode}
\usepackage{algorithmicx,algorithm}

\usepackage{algorithm}
\usepackage{algpseudocode}
\usepackage{amsmath}
\usepackage{pifont}
\usepackage{cite}
\usepackage{bm}
\usepackage{cleveref}
\usepackage{multicol}       
\usepackage{multirow}       
\usepackage{array}          
\usepackage{colortbl}
\usepackage{makecell}
\definecolor{crimson}{RGB}{192,0,0}         
\definecolor{navy}{RGB}{47,85,151}         
\usepackage{bbding}
\usepackage{graphicx}
\usepackage{booktabs}
\usepackage{algorithm}
\usepackage{algpseudocode}

\theoremstyle{plain}

\theoremstyle{plain}
\newtheorem{rem}{Remark}
\newtheorem{them}{Theorem}

\usepackage{amsmath}

\IEEEoverridecommandlockouts
\begin{document}

\title{Joint AP-UE Association and Precoding for SIM-Aided Cell-Free Massive MIMO Systems}

\author{Enyu Shi, Jiayi~Zhang,~\IEEEmembership{Senior Member,~IEEE}, Jiancheng~An, Guangyang Zhang, Ziheng Liu,\\ Chau Yuen,~\IEEEmembership{Fellow,~IEEE}, and Bo Ai,~\IEEEmembership{Fellow,~IEEE}
\thanks{E. Shi, J. Zhang, G. Zhang, Z. Liu, and B. Ai are with the School of Electronics and Information Engineering, Beijing Jiaotong University, Beijing 100044, P. R. China. (e-mail: \{enyushi, jiayizhang, guangyangzhang, zihengliu, boai\}@bjtu.edu.cn).}
\thanks{J. An and C. Yuen are with the School of Electrical and Electronics Engineering, Nanyang Technological University, Singapore 639798 (e-mail: jiancheng\_an@163.com, chau.yuen@ntu.edu.sg).}
}

\maketitle
\begin{abstract}
Cell-free (CF) massive multiple-input multiple-output (mMIMO) systems are emerging as promising alternatives to cellular networks, especially in ultra-dense environments. However, further capacity enhancement requires the deployment of more access points (APs), which will lead to high costs and high energy consumption. To address this issue, in this paper, we explore the integration of low-power, low-cost stacked intelligent metasurfaces (SIM) into CF mMIMO systems to enhance AP capabilities. The key point is that SIM performs precoding-related matrix operations in the wave domain. As a consequence, each AP antenna only needs to transmit data streams for a single user equipment (UE), eliminating the need for complex baseband digital precoding.
Then, we formulate the problem of joint AP-UE association and precoding at APs and SIMs to maximize the system sum rate. Due to the non-convexity and high complexity of the formulated problem, we propose a two-stage signal processing framework to solve it. In particular, in the first stage, we propose an AP antenna greedy association (AGA) algorithm to minimize UE interference. In the second stage, we introduce an alternating optimization (AO)-based algorithm that separates the joint power and wave-based precoding optimization problem into two distinct sub-problems: the complex quadratic transform method is used for AP antenna power control, and the projection gradient ascent (PGA) algorithm is employed to find suboptimal solutions for the SIM wave-based precoding.
Finally, the numerical results validate the effectiveness of the proposed framework and assess the performance enhancement achieved by the algorithm in comparison to various benchmark schemes. The results show that, with the same number of SIM meta-atoms, the proposed algorithm improves the sum rate by approximately 275\% compared to the benchmark scheme.
\end{abstract}

\begin{IEEEkeywords}
Stacked intelligent metasurface, cell-free massive MIMO, AP-UE association, power control, wave-based beamforming.
\end{IEEEkeywords}

\IEEEpeerreviewmaketitle

\section{Introduction}
\IEEEPARstart{T}{he} upcoming sixth-generation (6G) network is anticipated to play a pivotal role in numerous areas of future society, industry, and everyday life, necessitating extremely high communication standards regarding capacity, latency, reliability, and intelligence \cite{wang2023road}. The cell-free (CF) massive multiple-input multiple-output (mMIMO) system is proposed as a potential substitute for traditional cellular networks, offering notable benefits in ultra-dense environments due to its decentralized network architecture. In CF mMIMO systems, a large number of access points (APs), connected to a central processing unit (CPU) via fronthaul, are evenly dispersed throughout the service area. This arrangement enhances spatial diversity and minimizes the distance between transmitters and receivers \cite{ngo2017cell}. The APs work together to deliver communication services to user equipments (UEs) simultaneously, resulting in an enhanced quality of service.

Despite the performance enhancements brought by the introduction of CF mMIMO systems, the deployment of numerous access points (APs) has raised concerns about increased energy consumption and hardware costs. To address these issues, reconfigurable intelligent surfaces (RISs) have been proposed as an effective solution to boost CF mMIMO system performance while reducing energy usage and costs \cite{10556753,ma2022cooperative,wu2019intelligent,wang2024tutorial}. Extensive research has been conducted on integrating CF and RIS to fully exploit their combined benefits.
For instance, the authors in \cite{10167480} explored the uplink spectral efficiency (SE) of RIS-assisted CF mMIMO systems under electromagnetic interference (EMI) and proposed a fractional power control method to mitigate performance degradation. Additionally, \cite{zhang2021joint} introduced a joint RIS phase shift and AP precoding framework for a wideband RIS-assisted CF network to maximize the sum rate, demonstrating that strategic RIS phase shift design can significantly enhance system throughput, with optimal performance achieved by positioning RIS closer to APs. Also, the authors in \cite{papazafeiropoulos2024achievable} investigated the achievable rate of SIM-assisted holographic MIMO systems with an iterative optimization algorithm. Furthermore, the study in \cite{le2021energy} incorporated a RIS to reduce the energy consumption of CF mMIMO systems, aiming to maximize system energy efficiency (EE) through optimized RIS phase shifts.
Albeit these significant performance improvements, the dynamic nature of wireless environments necessitates frequent joint AP-RIS beamforming optimization, leading to increased signal processing complexity \cite{an2021low,10129196}. Additionally, the quasi-passive nature of single-layer RIS, along with hardware limitations and significant path loss, constrains its ability to implement advanced MIMO functionalities.

Fortunately, motivated by the rapid development of metasurface design, recent studies on intelligent surfaces have suggested using multi-layer metasurfaces for signal processing within the electromagnetic (EM) wave domain \cite{doi:10.1126/science.aat8084,liu2022programmable,nerini2024physically,liu2024stacked}. For instance, the research in \cite{liu2022programmable} introduced a programmable diffractive deep neural network structure utilizing a multilayer metasurface array, where each meta-atom acts as a reconfigurable artificial neuron. Inspired by this concept, the authors in \cite{10158690} proposed a novel stacked intelligent metasurface (SIM)-assisted MIMO transceiver. This framework involves layering multiple nearly passive, programmable metasurfaces to form a SIM configured similarly to an artificial neural network (ANN), significantly enhancing signal processing capabilities. They proposed that by deploying and configuring a pair of SIMs, each transceiver antenna only needs to transmit the corresponding data stream for a single UE, eliminating the need for complex multi-stream precoding, which significantly reduces the computational complexity of the transmitter.
Additionally, studies such as \cite{nadeem2023hybrid,papazafeiropoulos2024performance,wang2024multi} have leveraged SIM for receiver combining and transmit precoding in holographic MIMO communications, showing that SIM outperforms single-layer RIS in system performance. Further research by \cite{yao2024channel,an2024two} focused on channel and direction-of-arrival estimation to advance SIM development. As a result, SIM demonstrates great potential in facilitating the deployment of MIMO transceivers with advanced wave-domain transmit precoding and receive combining, while also significantly reducing RF energy consumption and hardware costs \cite{an2023stacked3,liu2024drl}.

Hence, empowering APs with SIM in CF mMIMO systems can leverage the strengths of both, reducing the complexity of hardware design and precoding at the AP, addressing the high power consumption issues of active antennas, and ultimately enhancing overall system performance \cite{9743355,hassan2024efficient}. For example, the authors in \cite{li2024stacked} investigated the uplink performance of the SIM for holographic MIMO-aided CF mMIMO systems. They proposed a layer-by-layer iterative optimization algorithm to design the hybrid beamforming at APs and SIMs. The results show that the introduction of SIM can significantly enhance the performance of the CF mMIMO system.
However, integrating CF with SIM introduces several pressing challenges that need to be addressed, including the design of joint precoding for AP power allocation and SIM phase shift, AP antenna and UE association \cite{masoumi2019performance}. 
First, we need to consider the collaboration between multiple APs and SIMs to achieve efficient signal transmission. Since the primary goal of SIM is to alleviate the load on APs, the signals transmitted by AP antennas need to be beamformed in the SIM beam domain before being transmitted to the UEs. Therefore, a joint design of the signal processing mechanisms for both APs and SIMs is required to enable their coordinated operation, thereby achieving interference mitigation between UEs and enhancing system performance \cite{li2024stacked,an2023stacked2}. This introduces a critical challenge: \textit{the joint precoding design of AP power allocation and SIM phase shifts}.
More crucially, the primary purpose of SIM is to alleviate the load at the AP, meaning that each AP antenna is only required to transmit the data stream for a single UE. Thus, AP antennas merely need to manage transmission power, eliminating the need for complex hardware infrastructure and multi-user precoding designs. Generally, in CF mMIMO systems, it is customary to deploy numerous low-complexity, basic APs with fewer antennas than the number of UEs in the system. This configuration does not guarantee that every SIM-aided AP can serve all UEs, introducing a new challenge: \textit{the association of AP antennas and UEs}, which has not been explored in CF mMIMO systems. Previous studies all considered scenarios where each antenna serves all UEs or where the number of UEs in the system is equal to or less than the number of antennas at the AP/base station (BS) \cite{an2023stacked3,li2024stacked,lin2024stacked,papazafeiropoulos2024performance}. 

Motivated by the observation above, we propose the concept of SIM-aided CF mMIMO systems. The key idea is to replace traditional APs with SIM-aided APs, thereby reducing the hardware design complexity and precoding computational demands of APs. We propose a scenario design for a joint AP antenna-UE association and precoding framework based on an alternating optimization (AO) algorithm to maximize the sum rate of the considered system. The contributions of this work are delineated as follows:

\begin{itemize}
\item We introduce the concept of integrating SIM into the CF mMIMO, referred to as SIM-aided CF mMIMO, to further enhance system performance. In the proposed system, each AP is equipped with a SIM, which significantly reduces the hardware and computational complexity of the AP. Remarkably, the transmit precoding is implemented automatically as the EM waves propagate through multiple metasurfaces in the SIM. Therefore, this configuration requires each antenna at the AP to transmit one data stream dedicated to a single UE, supplemented by straightforward power control, thus eliminating the need for complex baseband digital precoding.

\item To fully integrate SIM with CF mMIMO systems, we formulate a joint AP antenna association and precoding design problem at the APs and SIMs to maximize the sum rate for all UEs. To the best of our knowledge, this represents the first work considering the association between AP antennas and UEs in SIM-aided CF mMIMO systems. The formulated joint optimization problem is a mixed integer non-linear programming (MINLP) problem, where the AP antenna association matrix, transmit power allocation coefficients, and the phase shifts are deeply coupled within the non-convex objective function, making it challenging to obtain the optimal solution.

\item To address this challenge, we propose a two-stage signal processing framework. Specifically, in the first stage, we introduce an AP antenna greedy association (AGA) algorithm based on large-scale channel state information (CSI) to minimize UE interference. In the second stage, we propose an AO-based algorithm to transform the joint power allocation and wave-based precoding optimization problem into two sub-problems to be solved separately. For the antenna power allocation, we adopt the complex quadratic transform method. For the SIM wave-based precoding, we propose the projection gradient ascent (PGA) algorithm to obtain suboptimal solutions. Then, we analyze the complexity of the proposed two-stage signal processing framework.

\item Simulation results demonstrate that SIMs can significantly improve the CF mMIMO system capacity. Meanwhile, the results indicate that AP antenna-UE association is crucial in SIM-aided CF mMIMO systems, and a well-planned AP-UE association strategy can significantly enhance system performance. Additionally, the proposed two-stage optimization framework achieves a performance improvement of 275\% compared to the random phases and power optimization scheme.
\end{itemize}

The remainder of this paper is structured as follows. Section \uppercase\expandafter{\romannumeral2} outlines the SIM-aided CF mMIMO system model, including the channel model, AP-UE association, downlink data transmission, and problem formulation. Next, Section \uppercase\expandafter{\romannumeral3} proposes the two-stage joint AP antenna-UE association and precoding framework, consisting of the first stage of AP antenna-UE association and the second stage of AO-based joint phase shift design and power allocation. Then, numerical results and discussion are provided in Section \uppercase\expandafter{\romannumeral4}. Finally, Section \uppercase\expandafter{\romannumeral5} concludes this paper.

\textbf{Notation:} The column vectors and matrices are denoted by boldface lowercase letters $\mathbf{x}$ and boldface uppercase letters $\mathbf{X}$, respectively. The superscripts $\mathbf{x}^{\rm{H}}$, $x^\mathrm{T}$, and $x^\mathrm{*}$ are adopted to represent conjugate, transpose, and conjugate transpose, respectively. The $\triangleq$, $\left\|  \cdot  \right\|$, and $\left\lfloor  \cdot  \right\rfloor $ denote the definitions, the Euclidean norm, and the truncated argument, respectively. ${\rm{tr}}\left(  \cdot  \right)$, $\mathbb{E}\left\{  \cdot  \right\}$, and ${\rm{Cov}}\left\{  \cdot  \right\}$ denote the trace, expectation, and covariance operators, respectively. We exploit ${\text{diag}}\left( {{a_1}, \cdots ,{a_n}} \right)$ to express a diagonal matrix. 
The circularly symmetric complex Gaussian random variable $x$ with mean $0$ and variance $\sigma^2$ is denoted by $x \sim \mathcal{C}\mathcal{N}\left( {0,{\sigma^2}} \right)$. Then, $\nabla$ denotes the gradient operation. $\mathbb{B}^n$, $\mathbb{Z}^n$, $\mathbb{R}^n$, and $\mathbb{C}^n$ represent the $n$-dimensional spaces of binary, integer, real, and complex numbers, respectively. Finally, the $N \times N$ zero matrix and identity matrix are denoted by $\mathbf{0}_{N}$ and $\mathbf{I}_{N}$, respectively.


\begin{figure*}[t]
\centering
\includegraphics[scale=1]{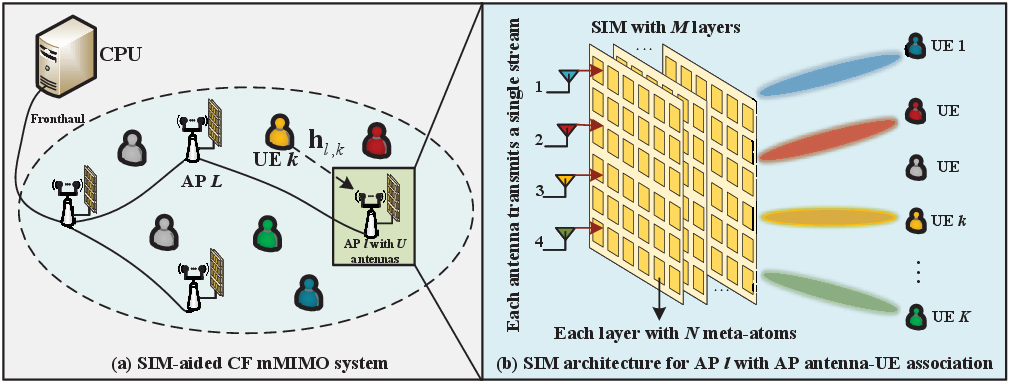}\vspace{-0.2cm}
\caption{Illustration of SIM-aided CF mMIMO systems.}\label{system_model}\vspace{-0.4cm}
\end{figure*}

\vspace{-0.1cm}

\section{System Model of the SIM-aided CF mMIMO Network}\label{se:model}
As shown in Fig.~\ref{system_model}, we consider a SIM-aided CF mMIMO system that consists of $L$ SIM-aided APs, a CPU, and $K$ UEs. We assume that each AP and UE is equipped with $U$ antennas and a single antenna, respectively. Specifically, we assume that each SIM is equipped with the same structure, which has $M$ metasurface layers and $N$ meta-atoms in each layer. Let ${\cal L} = \{ 1, \ldots ,L\}$, ${\cal K} = \{ 1, \ldots ,K\}$, ${\cal U} = \{ 1, \ldots ,U\}$, ${\cal M} = \{ 1, \ldots ,M\}$, and ${\cal N} = \{ 1, \ldots ,N\}$ denote the index sets of APs, UEs, AP antennas, SIM metasurface layers, and meta-atoms per layer, respectively. Furthermore, the SIM is connected to an intelligent controller at the AP, capable of applying a distinct and tunable phase shift to the EM waves passing through each meta-atom \cite{an2023stacked2}. In this scenario, there is a one-to-one binding relationship between SIM and AP, which means that the number of SIMs is equal to the number of APs, denoted as $L$. All APs are connected to the CPU via fronthaul links which can send the AP data to the CPU for centralized signal processing. In line with conventional CF mMIMO systems, it is postulated that the time division duplex (TDD) protocol is employed for the considered SIM-aided CF mMIMO system. 

\subsection{Channel Model}
We use ${e^{j\varphi _{l,m}^n}},\forall l \in {\cal L},\forall m \in {\cal M},\forall n \in {\cal N}$ with  $\varphi _{l,m}^n \in \left[ {0,2\pi } \right)$ denote the $n$-th meta-atom's phase shift in the $m$-th metasurface layer at the $l$-th SIM. Hence, the diagonal phase shift matrix ${{\bf{\Phi }}_{l,m}}$ for the $m$-th metasurface layer at the $l$-th SIM can be denoted as ${{\bf{\Phi }}_{l,m}} = {\rm{diag}}( {{e^{j\varphi _{l,m}^1}},{e^{j\varphi _{l,m}^2}}, \ldots ,{e^{j\varphi _{l,m}^N}}} ) \in \mathbb{C} {^{N \times N}},\forall l \in {\cal L},m \in {\cal M}$. Furthermore, let ${\bf{W}}_{l,1}^{} = {[ {{\bf{w}}_{l,1}^1, \ldots ,{\bf{w}}_{l,1}^{U}} ]} \in \mathbb{C} {^{N \times U}}$ denote the transmission vector from the $l$-th AP to the first metasurface layer of the SIM, where ${{\bf{w}}_{l,1}^{{u}}} \in \mathbb{C} {^{N}}$ denotes the $u$-th antenna of AP $l$ to the first metasurface layer within the SIM. Let ${{\bf{W}}_{l,m}} \in \mathbb{C} {^{N \times N}},\forall m \ne 1,m \in {\cal M},l \in {\cal L}$ denote the transmission matrix from the $(m-1)$-th to the $m$-th metasurface layer of SIM $l$. According to the Rayleigh-Sommerfeld diffraction theory introduced by \cite{lin2024stacked}, the $\left( {n,n'} \right)$-th element is expressed as

\begin{align}\label{w}
w_{l,m}^{n,n'} = \frac{{{d_x}{d_y}\cos \chi _{l,m}^{n,n'}}}{{d_{l,m}^{n,n'}}}\left( {\frac{1}{{2\pi d_{l,m}^{n,n'}}} - j\frac{1}{\lambda }} \right){e^{j2\pi \frac{{d_{l,m}^{n,n'}}}{\lambda }}},
\end{align}
where $\lambda$ represents the carrier wavelength, $d_{l,m}^{n,n'}$ indicates the corresponding transmission distance, $d_{x} \times d_{y}$ indicates the size of each SIM meta-atom, and ${\chi _{l,m}^{n,n'}}$ represents the angle between the propagation direction and the normal direction of the $(m-1)$-th SIM metasurface layer of SIM $l$. Similarly, the $n$-th element $w_{l,1,n}^u$ of ${{\bf{w}}_{l,1}^{{u}}}$ can be obtained from \eqref{w}.
Hence, the wave-based beamforming matrix ${\mathbf{G}_l} \in \mathbb{C}{^{N \times N}}$ of AP $l$, enabled by SIM, is obtained as 
\begin{align}\label{G_l}
{{\bf{G}}_l} = {{\bf{\Phi }}_{l,M}}{{\bf{W}}_{l,M}}{{\bf{\Phi }}_{l,M - 1}}{{\bf{W}}_{l,M - 1}} \ldots {{\bf{\Phi }}_{l,2}}{{\bf{W}}_{l,2}}{{\bf{\Phi }}_{l,1}}.
\end{align}

Then, we consider a quasi-static flat-fading channel model. 
Let ${\bf{h}}_{{\rm{SI}}{{\rm{M}}_l},k} \in \mathbb{C}{^{N \times 1}}$ denote the direct channel between the last metasurface layer of the SIM $l$ to UE $k$. Specifically, we assume that ${\bf{h}}_{{\rm{SI}}{{\rm{M}}_l},k}$ is characterized as the spatially correlated Rayleigh fading channel with  ${\bf{h}}_{{\rm{SI}}{{\rm{M}}_l},k} \sim {\cal C}{\cal N}\left( {0,{{\bf{R}}_{{\rm{SI}}{{\rm{M}}_l},k}}} \right)$. ${{\bf{R}}_{{\rm{SI}}{{\rm{M}}_l},k}} = {\beta _{{\rm{SI}}{{\rm{M}}_l},k}}{\bf{R}} \in \mathbb{C}{^{N \times N}}$ where ${\beta _{{\rm{SI}}{{\rm{M}}_l},k}}$ denotes the distance-dependent path loss between UE $k$ and $l$-th SIM, and the covariance matrix ${\bf{R}} \in \mathbb{C}{^ {N \times N}}$ describes the spatial correlation among different meta-atoms of the output metasurface layer within the SIM. Considering an isotropic scattering environment with multipath components uniformly distributed, the $\left( {n,n'} \right)$-th element of $\mathbf{R}$ is ${{\bf{R}}_{n,n'}} = {\rm{sinc}}\left( {2{{{d_{n,n'}}} \mathord{\left/
 {\vphantom {{{d_{n,n'}}} \lambda }} \right.
 \kern-\nulldelimiterspace} \lambda }} \right)$ \cite{bjornson2020rayleigh}, where $d_{n,n'}$ denotes the spacing distance between the meta-atoms and ${\rm{sinc}}\left( x \right) = {{\sin \left( {\pi x} \right)} \mathord{\left/
 {\vphantom {{\sin \left( {\pi x} \right)} {\left( {\pi x} \right)}}} \right.
 \kern-\nulldelimiterspace} {\left( {\pi x} \right)}}$ denotes the normalized sinc function. Then, the total channel ${{\bf{h}}_{l,k}} \in \mathbb{C}{^{U \times 1}}$ between AP $l$ and UE $k$ can be denoted as
\begin{align}\label{h_lk}
{{\bf{h}}_{l,k}} &= {\bf{W}}_{l,1}^{\rm{H}}{\bf{G}}_l^{\rm{H}}{\bf{h}}_{{\rm{SI}}{{\rm{M}}_l},k}.
\end{align}
\begin{rem}
We observe that in \eqref{h_lk}, the SIM modifies the channel state by adjusting the phase shifts within $\mathbf{G}_l$. Unlike the uncontrollable direct channels in traditional MIMO systems, this channel is affected by signals aggregated from multiple metasurface layers of the SIM, which provides the opportunity to mitigate the inter-user interference as the EM waves propagate through it.
\end{rem}

\subsection{AP-UE Association}
In CF mMIMO systems, performance enhancement is achieved by deploying more APs than the number of UEs. For considerations of low cost and low power, typically, each AP has fewer antennas than the number of UEs (i.e., $U<K$) \cite{bjornson2019making, ngo2017cell,liu2024graph}. In traditional CF mMIMO systems, each antenna at the AP can transmit multiple data streams for multiple UEs through baseband digital precoding, thereby serving a large number of UEs simultaneously. However, the cost of this approach is the need for complex precoding design, which significantly increases the computational complexity and hardware design requirements of the signal processing at the AP \cite{zhang2021joint}. Fortunately, the introduction of SIM can significantly reduce the computational burden on APs by unloading complex precoding signal processing from the AP antennas to the multi-layer metasurfaces of SIM. As mentioned in \cite{an2023stacked,10158690}, each antenna of a SIM-aided AP only needs to perform simple power control and transmit the corresponding data stream for a single UE. Therefore, in SIM-aided CF systems, it is essential to consider the issue of AP antenna and UE association. 

We assume that all $U$ antennas of AP $l$ transmit a single data stream for an individual UE and do not require any precoding design, only needing to control the transmission power ${{\bf{P}}_l} = {\rm{diag}}\left( {\left[ {\sqrt {{p_{l,1}}} , \ldots, \sqrt {{p_{l,u}}} ,\ldots,\sqrt {{p_{l,U}}} } \right]} \right) \in \mathbb{C}{^{U \times U}}$, where $p_{l,u}$ denotes the transmission power of $u$-th antenna at AP $l$. For the AP antenna-UE association, we use matrix $\mathbf{A}_l\in \mathbb{C}{^{U \times K}}$ to represent the association relationship between AP $l$ and all UEs, which can be written as 
\begin{align}\label{A_l}
{{\bf{A}}_l} = \left[ {\begin{array}{*{20}{c}}
{{a_{l,1,1}}}& \cdots &{{a_{l,1,K}}}\\
 \vdots & \ddots & \vdots \\
{{a_{l,U,1}}}& \cdots &{{a_{l,U,K}}}
\end{array}} \right] ,
\end{align}
where ${a_{l,u,k}} \in \{ {0,1} \},\forall l \in L,\forall u \in U,\forall k \in K$.  For ease of subsequent description, we denote each column and each row of matrix $\mathbf{A}_l$ as ${{\bf{a}}_{l,k}} = {\left[ {{a_{l,1,k}}, \ldots ,{a_{l, U, k}}} \right]^{\rm{T}}} \in \mathbb{C}{^{U \times 1}}$ and ${{\bf{a}}_{l,u}} = \left[ {{a_{l,u,1}}, \ldots,{a_{l,u,K}}} \right] \in \mathbb{C}{^{1 \times K}}$, respectively. 

\begin{rem}
    Note that, unlike traditional CF mMIMO systems, the power coefficient diagonal matrix $\mathbf{P}_l$ is only related to the number of antennas at AP $l$, which means that each antenna only needs to control its own transmission power without the complex baseband digital precoding. Besides, each row $\mathbf{a}_{l,u}$ of the association matrix ${\mathbf{A}_l}$ represents the UE selected by the $u$-th antenna of AP $l$. Since each antenna transmits a single data stream for one UE, each row $\mathbf{a}_{l,u}$ has exactly one entry that is 1, with all others being 0. However, multiple antennas on one AP can simultaneously transmit required data to the same UE, which means that each column $\mathbf{a}_{l,k}$ can have multiple entries that are 1. 
    Note that if each UE must be associated with at least one AP antenna, an additional condition must be achieved, i.e., $\sum\nolimits_{l = 1}^L {\sum\nolimits_{u = 1}^U {{a_{l,u,k}}} }  \ge 1,\forall k \in K$. 
    By adjusting the position of the 1 in ${\mathbf{A}_l}$, the association of AP antennas and UEs can be accomplished.
    
\end{rem}

\subsection{Downlink Data Transmission}
The downlink transmitted signal ${{\bf{x}}_l} \in \mathbb{C}{^{U \times 1}}$ at AP $l$ can be written as
\begin{align}\label{x_l}
    {{\bf{x}}_l} = {{\bf{P}}_l}{{\bf{A}}_l}{{\bf{s}}},
\end{align}
where ${{\bf{s}}} = {\left[ {{s_1}, \ldots ,{s_K}} \right]^{\rm{T}}} \in \mathbb{C}{^{K \times 1}}$ represents the transmission symbols of all $K$ UEs. 
Then, the received signal ${y_k} \in \mathbb{C}$ at UE $k$ can be expressed as 
\begin{align}\label{y_k}
{y_k} &= \sum\limits_{l = 1}^L {{\bf{h}}_{l,k}^{\rm{H}} {\bf{x}}_l}  + {{{n}}_k} = \sum\limits_{l = 1}^L {{\bf{h}}_{{{\rm{SIM}}_l},k}^{\rm{H}}{{\bf{G}}_l}{{\bf{W}}_{l,1}}{{\bf{P}}_l}{{\bf{A}}_l}{\bf{s}}}  + {n_k}\notag\\
 &= \sum\limits_{l = 1}^L {{\bf{h}}_{{{\rm{SIM}}_l},k}^{\rm{H}}{{\bf{G}}_l}{{\bf{W}}_{l,1}}{{\bf{P}}_l}{{\bf{a}}_{l,k}}{s_k}}  \notag\\
 &+ \mathop \sum \limits_{j = 1,j \ne k}^K \mathop \sum \limits_{l = 1}^L {\bf{h}}_{{{\rm{SIM}}_l},k}^{\rm{H}}{{\bf{G}}_l}{{\bf{W}}_{l,1}}{{\bf{P}}_l}{{\bf{a}}_{l,j}}{s_j} + {n_k},
\end{align}
where ${n_k} \sim {\cal CN}(0,\sigma_k^2)$ denotes the additive noise at UE $k$. Note that the first term on the right-hand side of \eqref{y_k} is the desired signal to UE $k$, while the second term denotes the interference from other UEs.

\subsection{Problem Fomulation}
Based on the system model above, we consider maximizing the sum rate of the proposed SIM-aided CF mMIMO system by designing the AP-UE association matrix, transmit power, and SIM phase shifts. At first, the signal-to-interference-plus-noise ratio (SINR) for the transmitted symbol $s_k$ at UE $k$ can be calculated by using \eqref{y_k} as
\begin{align}\label{gamma_k}
{\gamma _k} = \frac{{{{\left| {\sum\limits_{l = 1}^L {{\bf{h}}_{{\rm{SI}}{{\rm{M}}_l},k}^{\rm{H}}{{\bf{G}}_l}{{\bf{W}}_{l,1}}{{\bf{P}}_l}{{\bf{a}}_{l,k}}} } \right|}^2}}}{{\mathop \sum \limits_{j = 1,j \ne k}^K {{\left| {\sum\limits_{l = 1}^L {{\bf{h}}_{{\rm{SI}}{{\rm{M}}_l},k}^{\rm{H}}{{\bf{G}}_l}{{\bf{W}}_{l,1}}{{\bf{P}}_l}{{\bf{a}}_{l,j}}} } \right|}^2} + \sigma _k^2}}.
\end{align}
Thereby, the sum rate of all UEs is given by 
\begin{align}\label{R_sum}
{R_{\rm{sum}}} = \sum\limits_{k = 1}^K {{{\log }_2}\left( {1 + {\gamma _k}} \right)}.
\end{align}
Finally, the sum rate maximization optimization problem can be originally formulated as
\begin{subequations}\label{P_0}
  \begin{align}
&{{\cal P}^0}:\mathop {\max }\limits_{{p_{l,u}},\varphi _{l,m}^n,{a_{l,u,k}}} {R_{\rm{sum}}} = \sum\limits_{k = 1}^K {{\rm{lo}}{{\rm{g}}_2}(1 + {\gamma _k})} \label{0-1}\\
&\quad\quad{\rm{s}}.{\rm{t}}.\;\;{\rm{tr}}\left( {{{\left\| {{{\bf{P}}_l}} \right\|}^2}} \right) \le {P_{l,\max }}, \forall l \in {\cal L}, \label{0-2}\\
&\quad\quad\;\;\;\;\;\;\;{p_{l,u}} \ge 0, \forall l \in {\cal L},\forall u \in {\cal U}, \label{0-3}\\
&\quad\quad\;\;\;\;\;\;\;\varphi _{l,m}^n \in [0,2\pi ), \forall l \in {\cal L},\forall m \in {\cal M},\forall n \in {\cal N}, \label{0-4}\\
&\quad\quad\;\;\;\;\;\;\;{a_{l,u,k}} \in \{ {0,1} \}, \forall l \in {\cal L},\forall u \in {\cal U},\forall k \in {\cal K}, \label{0-5}\\
&\quad\quad\;\;\;\;\;\;\;\sum\limits_{k = 1}^K {{a_{l,u,k}} = 1}, \forall l \in {\cal L},\forall u \in {\cal U},\forall k \in {\cal K}, \label{0-6}\\
&\quad\quad\;\;\;\;\;\;\;\sum\nolimits_{l = 1}^L \!\!{\sum\nolimits_{u = 1}^U \!\!{{a_{l,u,k}}} }  \!\ge\! 1, \forall l \!\in\! {\cal L},\forall u \!\in\! {\cal U},\forall k \!\in\! \cal K,\label{0-7}
  \end{align}
\end{subequations}
where ${P_{l,\max }}$ denotes the maximum transmission power of each AP. It is clear that \eqref{0-2} and \eqref{0-3} guarantee the transmit power constraints of APs. Then, constraint \eqref{0-4} accounts for the limitation on phase shift for individual transmission meta-atoms of the SIM, representing a unit-modulus constraint, as evident from its nature.
Also, constraints \eqref{0-5}, \eqref{0-6}, and \eqref{0-7} represent the restrictions on AP-UE association, defining that each antenna can only select one UE and each UE must be associated by at least one antenna.

Note that the complex objective function \eqref{0-1} is non-convex, and the constant modulus constraint on the SIM transmission coefficients \eqref{0-4} is also non-convex, along with a zero-one discrete constraint \eqref{0-5} in the association matrix. Additionally, the SIM phase shifts, AP power coefficients, and antenna association coefficients are highly intertwined in the objective function, making it non-trivial to obtain the optimal solution for problem \eqref{P_0}. As such, in Section \ref{AO}, we will provide an efficient algorithm framework for solving the complex problem.

\section{Proposed Two-stage AP-UE Association and Precoding Framework} \label{AO}
In this section, a joint AP-UE association and precoding framework is proposed to solve the sum rate maximization optimization problem $\mathcal{P}^0$ in \eqref{P_0}. Specifically, an overview of the proposed framework is first provided in Subsection III-A, where the problem $\mathcal{P}^0$ is divided into three subproblems. Then, the detailed algorithms to solve the three subproblems are given in Sections \ref{A_selection}, \ref{power control}, and \ref{SIM phase shift}, respectively.

\subsection{Overview of the Proposed Framework}\label{Overview}
As a prerequisite for the proposed framework design, we assume that the global CSI of the whole SIM-aided CF mMIMO system can be fully acquired by the CPU via the fronthaul \cite{nadeem2023hybrid,yao2024channel}. Then, based on fully known CSI, joint AP-UE association, power control, and phase shifts at AP antennas and SIMs will be further designed. 

\begin{figure}[t]
\centering
\includegraphics[scale=0.6]{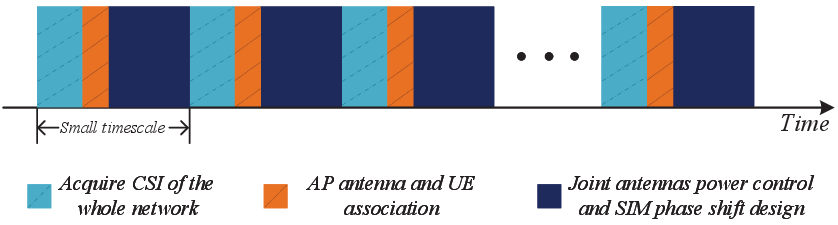}\vspace{-0.2cm}
\caption{Illustration of the transmission protocol.}\label{AO_framework}\vspace{-0.2cm}
\end{figure}

However, in SIM-aided CF mMIMO systems, before transmitting downlink data, it is necessary to complete the matching and access between UEs and APs, which requires addressing the issue of AP antenna and UE association first. Therefore, as shown in Fig.~\ref{AO_framework}, we divide the framework into three sequential sub-stages: CSI acquisition, AP antenna-UE association, and joint precoding optimization. In the first step, the association matrix ${\mathbf{A}}_l$ is designed based on the large-scale CSI between UEs and APs to facilitate one-to-one data transmissions from AP antennas to UEs. In the subsequent step, leveraging the established association matrix between AP antennas and UEs, the transmission power of the AP antennas and the phase shift matrix of the SIM are jointly optimized with AO. Finally, we address the optimization problem of maximizing the sum rate of the SIM-aided CF mMIMO system.

\subsection{AP Antenna and UE Association Design} \label{A_selection}
To facilitate the organic integration of SIM and CF systems and fully leverage the capabilities of SIM, thereby reducing the computational and hardware design complexity at APs, it is essential to assume that each antenna transmits only one UE data stream. Since the AP antenna and UE association occurs before data transmission, power control and phase optimization are not considered at this stage. 
Hence, a heuristic association strategy is developed by employing the fact that UEs closer to the AP experience less signal attenuation and are more likely to achieve better performance. Therefore, by calculating the distance tensor $\mathbf{D} \in {\mathbb{C}^{U \times K \times L}}$ between UEs and AP antennas, we can simplify the AP antenna-UE association problem to solving the minimization of the sum of association distances ${D_{{\rm{assoc}}}} = {\sum\nolimits_l {\sum\nolimits_k {\sum\nolimits_u {{({\bf{A}}_l)_{u,k}} \cdot ({\bf{D}})} } } _{u,k,l}}$. 
Based on this, $\mathcal{P}_0$ can be simplified as follows: 
\begin{subequations}\label{P_1}
\begin{align}
&{{\cal P}^1}:\mathop {\min }\limits_{a_{l,u,k}}\quad D_{\rm{assoc}}  \label{1-1}\\
&\quad\quad{\rm{s}}.{\rm{t}}.\;\;{a_{l,u,k}} \in \{ {0,1} \}, \forall l \in {\cal L},\forall u \in {\cal U},\forall k \in {\cal K}, \label{1-2}\\
&\quad\quad\;\;\;\;\;\;\;\sum\limits_{k = 1}^K {{a_{l,u,k}} = 1}, \forall l \in {\cal L},\forall u \in {\cal U},\forall k \in {\cal K},\label{1-3}\\
&\quad\quad\;\;\;\;\;\;\;\sum\nolimits_{l = 1}^L {\sum\nolimits_{u = 1}^U \!{{a_{l,u,k}}} }  \!\ge\! 1,\forall l \in {\cal L},\forall u \in {\cal U},\forall k \in \cal K. \label{1-4}
\end{align}
\end{subequations}
Note that constraint \eqref{1-2} defines the problem as a zero-one optimization problem, constraint \eqref{1-3} ensures that each antenna of the AP serves only one UE, and constraint \eqref{1-4} ensures that each UE must be associated by at least one antenna. To solve this non-convex problem, we propose an AP antenna greedy association (AGA) algorithm based on large-scale CSI. 
As shown in Algorithm 1, the proposed algorithm operates through a structured two-step process designed to optimize the allocation of antennas to UEs in the SIM-aided CF mMIMO system, which can be summarized into two stages as follows:
\subsubsection{Initialization and Distance Calculation} 
The AP AGA algorithm begins by initializing the necessary parameters and structures. Specifically, it inputs $L$, $U$, $K$, and AP and UE positions. Then, it calculates the distances between each UE and AP antenna, storing these values in the distance matrix $\mathbf{D}$.

\subsubsection{Greedy Antenna-UE Association Process} 
In the first stage, a greedy approach is employed to associate UEs with antennas. The algorithm iteratively finds the minimum distance pairs in $\mathbf{D}$, associating each UE to the nearest unassociated antenna of an AP. This association is recorded in the matrix $\mathbf{A}_l$ as
\begin{align}
{a_{l,u,k}} \!=\! 1 \Rightarrow {\bf{A}}_l\left( {u,k} \right) = 1,\,\forall l \in {\cal L},\forall u \in {\cal U},\forall k \in {\cal K}.
\end{align}
If all antennas of an AP are already associated, the algorithm removes that AP from the distance matrix to avoid reusing it. Based on that, we ensure that all UEs have at least one antenna associated with them as \eqref{1-4}.

\subsubsection{Final Antenna Allocation}
After completing the greedy association, the algorithm enters a while loop to ensure all antennas are associated. During this stage, the UEs are sorted based on their distances from each AP, and the remaining unassociated $u'$ antennas are assigned to the nearest $u'$ UEs. This process continues until all antennas at the AP side are fully utilized. This step ensures that all AP antennas have UEs to serve and will not remain idle. The final output is the matrix $\mathbf{A}$, representing the optimized AP antenna and UE associations.

\begin{rem}
    Note that the introduction of SIM, due to its advantage of simplifying data processing requirements at the AP, results in each antenna transmitting a single data stream. Consequently, this leads to the issue of antenna and UE association in the SIM-aided CF system, which is distinct from all AP antenna and UE association issues in traditional CF systems. This is a particularly critical problem in unleashing the full potential of SIM in CF mMIMO. 
    The proposed AGA is pivotal in enhancing the efficiency and performance of SIM-aided CF systems. By ensuring that every UE receives service and that all antennas are optimally utilized, AGA addresses critical challenges in resource allocation. Its greedy approach minimizes the distance between APs and UEs, leading to improved signal quality and reduced latency. Furthermore, its computational efficiency makes it suitable for real-time implementation in dynamic network environments, thereby significantly improving overall system performance and UE satisfaction. 
\end{rem}

\begin{algorithm}[t]
\caption{AP Antenna Greedy Association Algorithm for Solving \eqref{P_1} }
\begin{algorithmic}[1]
\State \textbf{Input:} $L$, $U$, $K$, AP and UE positions.
\State \textbf{Initialize:} ${\mathbf{A}_l} \gets \mathbf{0}_{U \times K}$.
\State Calculate the UE-AP distance pairs to obtain the distance matrix $\mathbf{D}$.
\For{$k = 1$ to $K$}
    \State Find the minimum value in $\mathbf{D}$, obtain the correspond-
    \Statex \hspace{\algorithmicindent}ing UE-AP pair, and then select an unassociated 
    \Statex \hspace{\algorithmicindent}antenna $u$ of the AP to associate the UE $k$. 
    \State Add the corresponding $a_{l, u, k}$ as 1 to $\mathbf{A}_l$.
    \State If all $U$ antennas of an AP are already associated with 
    \Statex \hspace{\algorithmicindent}UEs, delete all data about this AP in the correspond-
    \Statex \hspace{\algorithmicindent}ing distance matrix to avoid reuse.
\EndFor
\State \textbf{end for}
\While{not all antennas are associated}
    \For{$l = 1$ to $L$}
        \State Sort the UEs based on their distances from each 
        \Statex \hspace{\algorithmicindent}\hspace{\algorithmicindent}AP.
        \State Assign $u'$ unassociated antennas in the AP to the 
        \Statex \hspace{\algorithmicindent}\hspace{\algorithmicindent}top $u'$ ranked UEs. 
    \EndFor
    \State \textbf{end for}
\EndWhile
\State \textbf{end while}
\State \textbf{Output:} The AP antenna and UE association matrix $\mathbf{A}_l$.
\end{algorithmic}
\end{algorithm}

\subsection{Joint AP Power and SIM Configurable Design}\label{Joint_AO}
In this section, we propose the joint AP and SIM design to solve the sum rate maximization problem. Specifically, AP-UE association is a prerequisite for joint precoding, which has been addressed in Section \ref{A_selection}. Therefore, the optimization problem $\mathcal{P}^0$ in \eqref{P_0} can be simplified to jointly optimize AP power allocation and SIM phase shift design, as follows
\begin{subequations}\label{P_2}
  \begin{align}
&{{\cal P}^2}:\mathop {\max }\limits_{{p_{l,u}},\varphi _{l,m}^n} \quad{R_{\rm{sum}}}  \label{2-1}\\
&\quad\quad{\rm{s}}.{\rm{t}}.\;\;{\rm{tr}}\left( {{{\left\| {{{\bf{P}}_l}} \right\|}^2}} \right) \le {P_{l,\max }}, \forall l \in {\cal L}, \label{2-2}\\
&\quad\quad\;\;\;\;\;\;\;{p_{l,u}} \ge 0, \forall l \in {\cal L},\forall u \in {\cal U}, \label{2-3}\\
&\quad\quad\;\;\;\;\;\;\;\varphi _{l,m}^n \in [0,2\pi ), \forall l \in {\cal L},\forall m \in {\cal M},\forall n \in {\cal N}, \label{2-4}
  \end{align}
\end{subequations}
where \eqref{2-2} and \eqref{2-3} represent the constraints on AP transmission power, while \eqref{2-4} specifies the constraints on the phase shift of each meta-atoms on the SIM. Due to the non-convex nature of the objective function in $\mathcal{P}^2$, and the coupling of power and phase constraints within the objective function, inspired by \cite{10158690}, we employ the concept of an AO-based algorithm to transform $\mathcal{P}^2$ into two sub-optimization problems, which are then solved separately.

\subsubsection{Fix SIM Phase Shifts and Solve AP Power Control}\label{power control}
Given fixed SIM meta-atom phase shifts $\varphi^n_{l,m}$, the sum rate maximization problem ${\cal{P}}^2$ in \eqref{P_2} can be reformulated as the transmission power control at AP antennas, yielding
\begin{subequations}\label{P_power}
  \begin{align}
&{{\cal P}_{\rm{power}}}:\mathop {\max }\limits_{{p_{l,u}}} \quad{R_{\rm{sum}}}  \label{power-1}\\
&\quad\quad\quad\quad{\rm{s}}.{\rm{t}}.\;\;{\rm{tr}}\left( {{{\left\| {{{\bf{P}}_l}} \right\|}^2}} \right) \le {P_{l,\max }}, \forall l \in {\cal L}, \label{power-2}\\
&\quad\quad\quad\quad\;\;\;\;\;\;\;{p_{l,u}} \ge 0, \forall l \in {\cal L},\forall u \in {\cal U}. \label{power-3}
  \end{align}
\end{subequations}
To facilitate solving $\cal{P}_{\rm{power}}$, we perform a simple transformation on $\gamma_k$ as
\begin{align}
{\gamma _k} = \frac{{{{\left| {{\bf{h}}_k^{\rm{H}}{\bf{Q}}{{\bf{A}}_k}{\bf{p}}} \right|}^2}}}{{\mathop \sum \limits_{j = 1,j \ne k}^K {{\left| {{\bf{h}}_k^{\rm{H}}{\bf{Q}}{{\bf{A}}_j}{\bf{p}}} \right|}^2} + \sigma _k^2}},\forall k \in {\cal K},
\end{align}
where ${\bf{h}}_k^{\rm{H}} = \left[ {{\bf{h}}_{{\rm{SI}}{{\rm{M}}_1},k}^{\rm{H}}, \ldots ,{\bf{h}}_{{\rm{SI}}{{\rm{M}}_L},k}^{\rm{H}}} \right] \in {\mathbb{C}^{1 \times NL}}$, ${\bf{Q}} = {\rm{diag}}\left( {\left[ {{{\bf{G}}_1}{{\bf{W}}_{1,1}}, \ldots ,{{\bf{G}}_L}{{\bf{W}}_{L,1}}} \right]} \right) \in {\mathbb{C}^{NL \times UL}}$, ${{\bf{A}}_k} = {\rm{diag}}\left( {\left[ {{\rm{diag}}\left( {{{\bf{a}}_{1,k}}} \right), \ldots ,{\rm{diag}}\left( {{{\bf{a}}_{L,k}}} \right)} \right]} \right) \in {\mathbb{C}^{UL \times UL}}$, and ${\bf{p}} = {\left[ {{\rm{diag}}\left( {{{\bf{P}}_1}} \right), \ldots ,{\rm{diag}}\left( {{{\bf{P}}_L}} \right)} \right]^{\rm{T}}} \in {\mathbb{C}^{UL \times 1}}$. Then, the problem \eqref{P_power} can be equivalent to 
\begin{subequations}\label{P_power1}
  \begin{align}
&{\cal P}_{{\rm{power}}}^1:\mathop {\max }\limits_{{\mathbf{p}}} \sum\limits_{k = 1}^K {{\rm{lo}}{{\rm{g}}_2}(1 \!+\! \frac{{{{\left| {{\bf{h}}_k^{\rm{H}}{\bf{Q}}{{\bf{A}}_k}{\bf{p}}} \right|}^2}}}{{\mathop \sum \limits_{j = 1,j \ne k}^K {{\left| {{\bf{h}}_k^{\rm{H}}{\bf{Q}}{{\bf{A}}_j}{\bf{p}}} \right|}^2} \!\!+\! \sigma _k^2}})} \label{power1-1}\\
&\quad\quad\quad\quad{\rm{s}}.{\rm{t}}.\;\;\;\;{\left\| {{{\bf{E}}_l}{\bf{p}}} \right\|^2} \le {P_{l,\max }},\quad \forall l \in {\cal L}, \label{power1-2}\\
&\quad\quad\quad\quad\quad\;\;\;\;\;\;\;{p_{l,u}} \ge 0,\quad \forall l \in {\cal L},\forall u \in {\cal U}, \label{power1-3}
  \end{align}
\end{subequations}
where ${{\bf{E}}_l} = {\rm{diag}}(\underbrace {0, \ldots ,0}_{\left( {l - 1} \right)U},\underbrace {1, \ldots ,1}_U,\underbrace {0, \ldots ,0}_{\left( {L - l} \right)U}) \in {\mathbb{C}^{UL \times UL}}$ is utilized to select the $U$ antennas corresponding to the $l$-th AP.
The objective function in the problem \eqref{P_power1} is non-convex due to its fractional form. To deal with it, we apply the quadratic transform method by introducing auxiliary variables $t_k, {\bar \gamma }_k \in \mathbb{R}$ to obtain an objective function in \eqref{P_power1} as follows \cite{shen2018fractional}
\begin{align}
&{R_{\rm{sum}}}\left( {{\bf{p}},{t_k}} \right) = \sum\limits_{k = 1}^K {{\rm{lo}}{{\rm{g}}_2}\left( {1 + 2{t_k}\sqrt {{{\left| {{\bf{h}}_k^H{\bf{Q}}{{\bf{A}}_k}{\bf{p}}} \right|}^2}} } \right.} \notag\\
&\left. { - t_k^2\left( {\mathop \sum \limits_{j = 1,j \ne k}^K {{\left| {{\bf{h}}_k^H{\bf{Q}}{{\bf{A}}_j}{\bf{p}}} \right|}^2} + \sigma _k^2} \right)} \right).
\end{align}
Therefore, the problem \eqref{P_power1} is equivalent to 
\begin{subequations}\label{P_power2}
  \begin{align}
&{\cal P}_{{\rm{power}}}^2:\;\;\mathop {\max }\limits_{{\bf{p}},{t_k},{{\bar \gamma }_k}} \;\;\sum\limits_{k = 1}^K {{\rm{lo}}{{\rm{g}}_2}\left( {1 + {{\bar \gamma }_k}} \right)} \label{power2-1}\\
& {\rm{s}}.{\rm{t}}.\;\;2{t_k}\sqrt {{{\left| {{\bf{h}}_k^{\rm{H}}{\bf{Q}}{{\bf{A}}_k}{\bf{p}}} \right|}^2}}  \!\ge\! t_k^2\!\left( {\mathop \sum \limits_{j = 1,j \ne k}^K \!\!{{\left| {{\bf{h}}_k^{\rm{H}}{\bf{Q}}{{\bf{A}}_j}{\bf{p}}} \right|}^2} \!\!+\! \sigma _k^2} \right) \!\!+\! {{\bar \gamma }_k}, \label{power2-2}\\
& \quad \quad {\left\| {{{\bf{E}}_l}{\bf{p}}} \right\|^2} \le {P_{l,\max }},\quad \forall l \in {\cal L}, \label{power2-3}\\
& \quad \quad {p_{l,u}} \ge 0,\quad \forall l \in {\cal L},\forall u \in {\cal U}. \label{power2-4}
  \end{align}
\end{subequations}
Problem \eqref{P_power2} becomes a solvable convex optimization problem, which can be solved by iterative optimization algorithm in two steps as follows:
\begin{enumerate}
\item[\textit{i)}] Fix auxiliary variable $t_k$, and then optimize the variables $\mathbf{p}$ and ${{{\bar \gamma }_k}}$.
\item[\textit{ii)}] Fix variables $\mathbf{p}$ and ${{{\bar \gamma }_k}}$, and then optimize auxiliary variable $t_k$.
\end{enumerate}
By iteratively repeating the two steps mentioned above, when the growth rate of the sum rate falls below the given threshold, the iteration stops, thereby achieving power optimization.

\subsubsection{Fix AP Power and Solve SIM Phase Shifts Design}\label{SIM phase shift}
In the case of given AP antennas transmission power $\mathbf{p}$, the equivalent sum rate maximization problem $\mathcal{P}^2$ in \eqref{P_2} can be reformulated as the following subproblem $\mathcal{P}_{\rm{SIM}}$ for the phase shift design of meta-atoms on SIMs:
\begin{subequations}\label{P_SIM}
  \begin{align}
&{{\cal P}_{\rm{SIM}}}:\mathop {\max }\limits_{{p_{l,u}},\varphi _{l,m}^n} {R_{\rm{sum}}} \label{SIM-1}\\
&\quad\quad{\rm{s}}.{\rm{t}}.\;\;\varphi _{l,m}^n \in [0,2\pi ), \forall l \in {\cal L},\forall m \in {\cal M},\forall n \in {\cal N}. \label{SIM-2}
  \end{align}
\end{subequations}
In general, finding the globally optimal phase shifts for problem $\mathcal{P}_{\rm{SIM}}$ in \eqref{P_SIM} is quite challenging. 
To tackle this, we propose a computationally efficient gradient ascent algorithm to update the phase shifts $\varphi^n_{l,m}$ iteratively until converging to the vicinity of a stationary point. The detailed steps of the PGA algorithm are as follows
\begin{enumerate}
\item[\textit{i)}] Initialize the phase shifts of all SIM meta-atoms $\varphi _{l,m}^n \in [0,2\pi ), \forall l \in {\cal L},\forall m \in {\cal M},\forall n \in {\cal N}$. Next, compute the sum rate by using \eqref{R_sum}.
\item[\textit{ii)}] Then, we derive the partial derivative values of $R_{\rm{sum}}$ with respect to all $\varphi _{l,m}^n$, which is described in detail in Theorem \ref{them}.
\begin{them}\label{them}
For $\forall l \in {\cal L},\forall m \in {\cal M},\forall n \in {\cal N}$, the partial derivative of sum rate $R_{\rm{sum}}$ with respect to $\varphi _{l,m}^n$ can be derived as 
\begin{align}\label{deta_R}
    \frac{{\partial {R_{{\rm{sum}}}}}}{{\partial \varphi _{l,m}^n}} \!=\! {\log _2}e \!\cdot \!\sum\limits_{k = 1}^K {{\delta _k}\left( \!\!{{{\left( {\chi _{l,m}^n} \right)}_{k,k}} \!\!-\! {\gamma _k}\!\!\!\!\!\sum\limits_{j = 1,j \ne k}^K \!\!\!\!\!{{{\left( {\chi _{l,m}^n} \right)}_{k,j}}} } \!\right)}, 
\end{align}
where ${{\delta _k}}$ is denoted by
\begin{align}\label{dalta}
   {\delta _k} = \frac{1}{{\sum\limits_{j = 1}^K {{{\left| {\sum\limits_{l = 1}^L {{\bf{h}}_{l,k}^{\rm{H}}{{\bf{G}}_l}{{\bf{W}}_{l,1}}{{\bf{P}}_l}{{\bf{a}}_{l,j}}} } \right|}^2} + \sigma _k^2} }},
\end{align}
and ${\left( {\chi _{l,m}^n} \right)_{k,j}}$ is denoted by \eqref{chi} at the top of the next page. Furthermore, in \eqref{chi}, ${{\bf{b}}_{l,m}^n}$ and ${{{\left( {{\bf{q}}_{l,m}^n} \right)}^{\rm{H}}}}$ denote the $n$-th column of the matrix ${{\bf{B}}_{l,m}} \in {\mathbb{C}^{N \times N}}$ and $n$-th row of the matrix ${{\bf{Q}}_{l,m}} \in {\mathbb{C}^{N \times N}}$, respectively. Specially, the expressions of ${{\bf{B}}_{l,m}}$ and ${{\bf{Q}}_{l,m}}$ are defined by 
\newcounter{mytempeqncnt}
\begin{figure*}[t!]
\normalsize
\setcounter{mytempeqncnt}{1}
\setcounter{equation}{20}
\begin{align}\label{chi}
{\left( {\chi _{l,m}^n} \right)_{k,j}} = \frac{{\partial {{\left| {\sum\limits_{l = 1}^L {{\bf{h}}_{l,k}^{\rm{H}}{{\bf{G}}_l}{{\bf{W}}_{l,1}}{{\bf{P}}_l}{{\bf{a}}_{l,k}}} } \right|}^2}}}{{\partial \varphi _{l,m}^n}} = 2 \cdot \Im \left[ {\left( {{e^{j\varphi _{l,m}^n}}{\bf{h}}_{l,k}^{\rm{H}}{\bf{b}}_{l,m}^n{{\left( {{\bf{q}}_{l,m}^n} \right)}^{\rm{H}}}{{\bf{W}}_{l,1}}{{\bf{P}}_l}{{\bf{a}}_{l,k}}} \right)\left( {\sum\limits_{l = 1}^L {{\bf{h}}_{l,k}^{\rm{H}}{{\bf{G}}_l}{{\bf{W}}_{l,1}}{{\bf{P}}_l}{{\bf{a}}_{l,k}}} } \right)} \right].
\end{align}
\setcounter{equation}{21}
\hrulefill
\end{figure*}
\begin{align}\label{B}
    {{\bf{B}}_{l,m}} \!\buildrel \Delta \over =\!\! \left\{\!\!\!\! {\begin{array}{*{20}{c}}
{{{\bf{\Phi }}_{l,M}}{{\bf{W}}_{l,M}} \ldots {{\bf{\Phi }}_{l,\left( {m + 1} \right)}}{{\bf{W}}_{l,\left( {m + 1} \right)}},}&\!\!\!\!{{\rm{if}}\,m \!\ne\! M,}\\
{{{\bf{I}}_N},}&\!\!\!\!{{\rm{if}}\,m \!=\! M,}
\end{array}} \right.
\end{align}
\begin{align}\label{Q}
{{\bf{Q}}_{l,m}} \buildrel \Delta \over = \left\{ {\begin{array}{*{20}{c}}
{{{\bf{W}}_{l,m}}{{\bf{\Phi }}_{l,\left( {m - 1} \right)}} \ldots {{\bf{W}}_{l,2}}{{\bf{\Phi }}_{l,1}},}&{{\rm{if}}\,m \ne 1,}\\
{{{\bf{I}}_N},}&{{\rm{if}}\,m = 1,}
\end{array}} \right.
\end{align}

\end{them}

\begin{IEEEproof}
    The proof is given in Appendix A.
\end{IEEEproof}

\item[\textit{iii)}] Based on calculating all the partial derivatives of $R_{\rm{sum}}$ with respect to ${\varphi_{l,m}^n}$ as described in \eqref{deta_R}, we subsequently update all the phase shifts ${\varphi_{l,m}^n}$ simultaneously by
\begin{align}
\setcounter{equation}{23}
    \varphi _{l,m}^n \leftarrow \varphi _{l,m}^n + \eta \frac{{\partial {R_{{\rm{sum}}}}}}{{\partial \varphi _{l,m}^n}},\forall l \in {\cal L},\forall m \in {\cal M},\forall n \in {\cal N},
\end{align}
where $\eta  > 0$ denotes the $\mathit{Armijo}$ step size, determined using the backtracking line search method \cite{papazafeiropoulos2021intelligent}.

\item[\textit{iv)}] Repeat Steps \textit{ii)} to \textit{iii)} iteratively until the fractional increase in the sum rate is less than a predetermined threshold. Finally, return the corresponding numerical values of all $\varphi _{l,m}^n$, which are the optimized phase shifts of the SIM meta-atoms.

\end{enumerate}

\begin{rem}
    Note that the convergence of the PGA algorithm to a local maximum is ensured because: (1) the sum rate $R_{\rm{sum}}$ is upper bounded as shown in \cite[Eq.~(17)]{an2023stacked2}, and (2) the sum rate $R_{\rm{sum}}$ is non-decreasing when an appropriate Armijo step size $\eta$ is selected at each iteration \cite{armijo1966minimization}. However, the quantization process at each iteration may affect performance. To improve the robustness of the PGA algorithm, multiple initializations are employed.
\end{rem}

As a result, Algorithm 2 summarizes the entire AO-based algorithm for solving problem $\mathcal{P}^2$. To sum up, after determining the association matrix $\mathbf{A}_l$ through the AP antenna greedy association algorithm, the optimization program outlined in Section \ref{Joint_AO} can be iteratively executed to achieve the suboptimal solution of $\mathcal{P}^0$.

\begin{algorithm}[t]
\caption{Proposed joint AP Power Control and SIM Phase Shift Framework for Solving \eqref{P_2} }
\begin{algorithmic}[1]
\State \textbf{Input:} $\mathbf{h}_{{\rm{SIM}},k}$, $\mathbf{W}_{l,m}$, $\mathbf{A}_k$, 
\State \textbf{Initialize:} $\varphi _{l,m}^n \in \left[ {0,2\pi } \right)$.
\For{$iter = 1$ to $iter_{\rm{max}}$}
    \State Update $\mathbf{p}$ with the modified iterative optimization
    \Statex \hspace{\algorithmicindent}\hspace{\algorithmicindent}\!\!\!\!\!\!\!\!\!algorithm.
    \State \textbf{Repeat}
           \State \quad \quad Optimize $\mathbf{p}$ and ${{{\bar \gamma }_k}}$.
           \State \quad \quad Optimize $t_k$.
    \State \textbf{Until} Convergence of objective function.
    \State Update $\varphi^n_{l,m}$ with iterative gradient ascent algorithm.
\State \textbf{Until} The sum rate $R_{\rm{sum}}$ in \eqref{P_2} achieves convergence.
\EndFor
\State \textbf{Output:} The AP power control coefficient $p_{l,u}$, SIM meta-atoms phase shift $\varphi^n_{l,m}$, and $R_{\rm{sum}}$.
\end{algorithmic}
\end{algorithm}

\subsection{Complexity Analysis}
The complexity of the proposed framework consists of two main parts: the complexity associated with the association matrix $\mathbf{A}_l$ in \eqref{P_1} and the complexity of the AO-based algorithm in \eqref{P_2}. Furthermore, the complexity of the AO-based algorithm in \eqref{P_2} is divided into two parts: the complexity of the AP antenna power control algorithm in \eqref{P_power2} and the complexity for optimizing the SIM phase shift using the PGA algorithm in \eqref{P_SIM}. For the AP antenna and UE association, in particular, the complexity of the AGA in Section \ref{A_selection} is $\mathcal{O}_{\rm{Sel}}=\mathcal{O}(KU^2 + LKU)$. For the AP antenna power optimization, the computational complexity is $\mathcal{O}_{\rm{Power}} = \mathcal{O}({K^2}{L^3}{U^3}I_{\rm{AP}}(4N+3))$ by solving the second-order cone programming, where $I_{\rm{AP}}$ denotes the number of iterations for achieving convergence. For the phase shift optimization, the complexity of the gradient ascent algorithm to determine the phase shifts of SIM is given by $\mathcal{O}_{\rm{Phase}} = \mathcal{O}(2{K^2}LMNI_{\rm{GA}}(4N+3))$, where $I_{\rm{GA}}$ denotes the number of iterations for achieving convergence of the gradient ascent algorithm. As a result, the total complexity of the proposed AP antenna-UE association and joint power and phase shifts optimization framework can be expressed as $\mathcal{O}_{\rm{total}} = \mathcal{O}_{\rm{Sel}} + I_{\rm{AO}}(\mathcal{O}_{\rm{Power}}+\mathcal{O}_{\rm{Phase}})$, where $I_{\rm{AO}}$ is the number of iteration for AO. Note that the overall complexity increases cubically with the total number of antennas $LU$ across all APs and quadratically with the total number of SIM meta-atoms $LMN$ and the increase in the number of UEs $K$.

\section{Numerical Results and Discussion}
\subsection{Simulation Setup}
This section presents numerical results and discussion to validate the accuracy of the derived analysis and to evaluate the performance of the SIM-aided CF mMIMO system.
We assume that the APs and UEs are uniformly distributed in the $200 \times 200 \:\rm{m}^2$ area. Furthermore, a SIM, which stacks multiple metasurfaces, is integrated into the AP to execute transmit beamforming in the EM wave domain. The height of the SIM-aided APs and UEs is 15 $\rm{m}$ and 1.65 $\rm{m}$, respectively. Furthermore, we assume the thickness of the SIM is $T_{\rm{SIM}} = 5\lambda$, ensuring that the spacing between two adjacent metasurfaces in the $M$-layer SIM is ${d_{{\rm{Layer}}}} = {{{T_{{\rm{SIM}}}}} \mathord{\left/
 {\vphantom {{{T_{{\rm{SIM}}}}} M}} \right.
 \kern-\nulldelimiterspace} M}$. Furthermore, we consider a square metasurface structure SIM with $N={N_x}{N_y}$ meta-atoms and $N_x = N_y$ where $N_x$ and $N_y$ denotes the number of meta-atoms along the $x$-axis and $y$-axis, respectively. Furthermore, we assume half-wavelength spacing between adjacent antennas/meta-atoms at the APs and metasurfaces. Then, the size of SIM meta-atom is $d = {d_x} = {d_y} = {\lambda  \mathord{\left/
 {\vphantom {\lambda  2}} \right.
 \kern-\nulldelimiterspace} 2}$. Moreover, we assume a half-wavelength spacing between meta-atoms on SIMs and antennas at the APs. The spatial attenuation coefficients $w^{n,n'}_{l,m}$ and $d^{n,n'}_{l,m}$ between neighboring metasurface layers in SIM and that from the transmit antenna array to the first metasurface layer are determined by \cite[Eq.~(25)]{an2023stacked2}. Also, the transmission distance $d_{u,n}$ between the $u$-th antenna and the $n$-th meta-atom on the first metasurface layer is determined by \cite[Eq.~(26)]{an2023stacked2}. We assume a correlated Rayleigh fading channel model, and the distance-dependent path loss is modeled as 
 \begin{align}
      {\beta _{lk}} = {C_0}{\left( {{{{d_{lk}}} \mathord{\left/
 {\vphantom {{{d_{lk}}} {{d_0}}}} \right.
 \kern-\nulldelimiterspace} {{d_0}}}} \right)^\varpi },\quad {d_{lk}} > {d_0},
 \end{align}
where $d_{lk}$ denotes the link distance between the $l$-th SIM-aided AP to the $k$-th UE. Then, ${C_0} = {\left( {{\lambda  \mathord{\left/
 {\vphantom {\lambda  {4\pi {d_0}}}} \right.
 \kern-\nulldelimiterspace} {4\pi {d_0}}}} \right)^2}$ denotes the free space path loss with the reference distance $d_0 = 1$ m \cite{rappaport2015wideband}, and $\varpi =3.5 $ denotes the path loss exponent. Besides, we consider a system operating at a carrier frequency of 28 GHz with a transmission bandwidth of 10 MHz, and an effective noise power spectral density of $-174$ dBm/Hz.
 Each UE transmits with maximize power $200\; \rm{mW}$. For the PGA algorithm, the maximum allowable number of iterations is set to 100, with the initial learning rate and decay parameter set to 0.1 and 0.5, respectively, unless otherwise specified. All the simulation results are obtained by averaging over 100 independent experiments.

For convenience of comparison, the following schemes are listed.
\subsubsection{\textbf{AGA/NUA Proposed-AO}}
The AP antenna adopts a greedy association algorithm in Algorithm 1 or the nearest UE association (NUA) algorithm. Then, we employ the proposed AO-based algorithm to address AP antenna power control and SIM phase shift design.
\subsubsection{\textbf{AGA/NUA SIM-Opt}}
The AP antenna adopts the greedy/nearest UE association algorithm. Then, we employ the PGA algorithm to address SIM phase shift design with equal power transmission of AP antennas.
\subsubsection{\textbf{AGA/NUA Power-Opt}}
The AP antenna adopts the greedy/nearest UE association algorithm. Then, we employ the proposed power control algorithm to address AP antenna power allocation with random phase shifts of SIM meta-atoms.
\subsubsection{\textbf{AGA/NUA RP-EP}}
The AP antenna adopts the greedy/nearest UE association algorithm. Then, we assume that adopting random phase shifts of SIM meta-atom and equal power allocation of AP antennas.

\subsection{Algorithm Convergence}
\begin{figure}[t]
\centering
\includegraphics[scale=0.55]{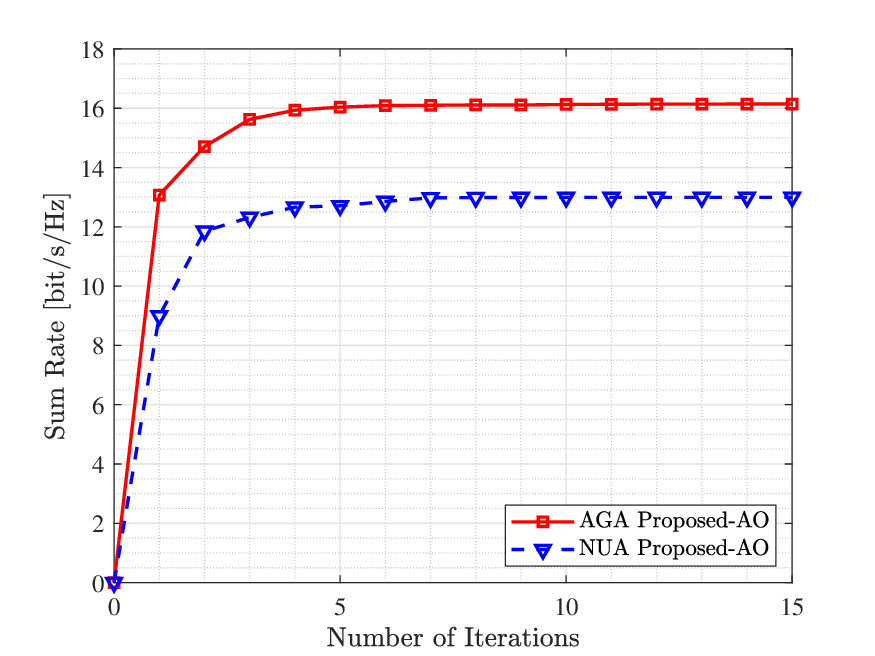}\vspace{-0.2cm}
\caption{Average sum rate against the numbers of iterations ($L = 6$, $U = 2$, $M = 2$, $N = 25$, $K = 4$, ${d_x} = {d_y} = {\lambda  \mathord{\left/
 {\vphantom {\lambda  2}} \right.
 \kern-\nulldelimiterspace} 2}$).}\label{Fig_iteration}\vspace{-0.2cm}
\end{figure}

In this subsection, we investigate the convergence rate of the proposed AP-based algorithms with different AP antenna-UE association schemes in SIM-aided CF mMIMO systems. As shown in Fig.~\ref{Fig_iteration}, it is clear that the proposed AO-based algorithm can achieve convergence with only a few outer iterations. This efficiency is due to the fact that power optimization relies more on large-scale CSI, while the optimization of SIM configuration predominantly affects phase shift changes. Thus, the changes in power and phase shifts in the optimization results have minimal impact on each other. Consequently, each component can reach its own convergence quickly in the iteration process, thereby accelerating the convergence of the overall AO process. Notably, after achieving convergence, introducing the AGA algorithm can enhance sum rate performance by 24\%, underscoring the effectiveness of our proposed AP-UE association algorithm.

\subsection{Impact of Parameters in CF mMIMO Systems}
\begin{figure}[t]
\centering
\includegraphics[scale=0.55]{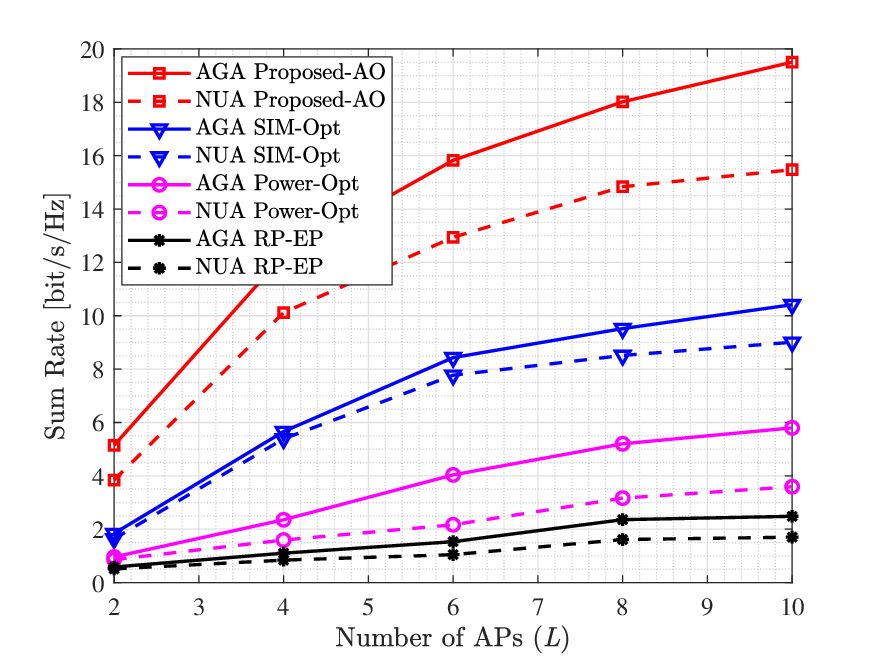}\vspace{-0.2cm}
\caption{Sum rate against different numbers of APs ($U = 2$, $M = 2$, $N = 25$, $K = 4$, ${d_x} = {d_y} = {\lambda  \mathord{\left/
 {\vphantom {\lambda  2}} \right.
 \kern-\nulldelimiterspace} 2}$).}\label{Fig_L}\vspace{-0.2cm}
\end{figure}

Fig.~\ref{Fig_L} illustrates the sum rate against different numbers of APs with different design schemes of the considered system. The results indicate that increasing the number of APs leads to an improvement in system performance. Moreover, as the number of APs $L$ increases, the performance improvement of the proposed AGA algorithm becomes more pronounced compared to the NUA algorithm. For example, at $L = 6$ and $L = 10$, the AGA algorithm shows performance improvements of 22\% and 26\%, respectively, compared to the NUA algorithm. This indicates that as the number of APs increases, an efficient AP-UE association algorithm becomes increasingly important. Also, when $L = 6$ and adopting the AGA algorithm, the sum rate using only SIM phase shifts optimization surpasses that using only antenna power control by 108\%. It is evident that in SIM-aided CF mMIMO systems, the impact of SIM phase shift design on system performance enhancement is far more significant than that of power control.

\begin{figure}[t]
\centering
\includegraphics[scale=0.55]{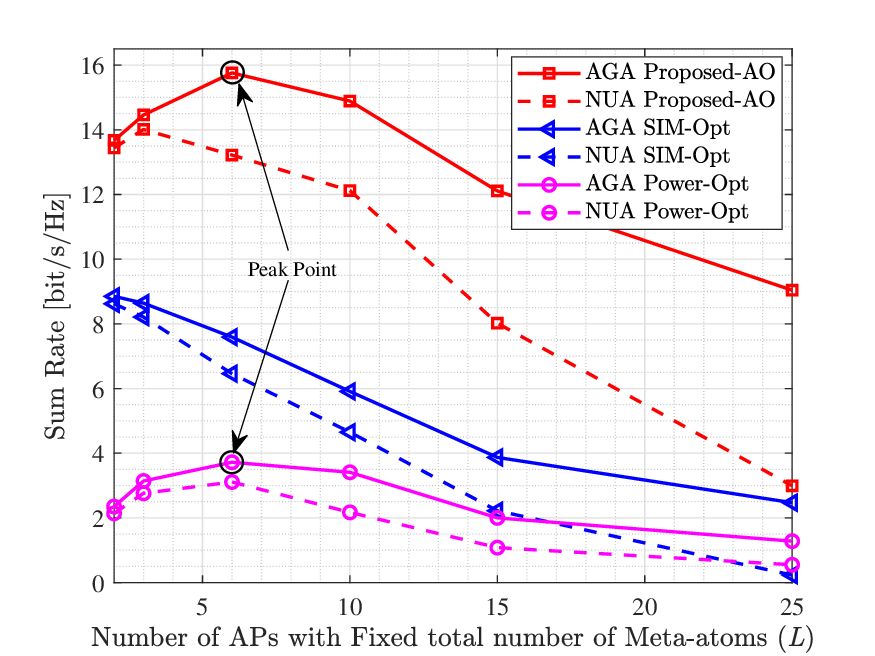}\vspace{-0.2cm}
\caption{Sum rate against different numbers of APs with the fixed total number of meta-atoms ($U = 2$, $M = 2$, $N_{\rm{total}} = 300$, $K = 4$, ${d_x} = {d_y} = {\lambda  \mathord{\left/
 {\vphantom {\lambda  2}} \right.
 \kern-\nulldelimiterspace} 2}$).}\label{Fig_N_fixed}\vspace{-0.2cm}
\end{figure}

Fig.~\ref{Fig_N_fixed} illustrates the sum rate against different numbers of APs with the fixed total number of SIM meta-atoms in the considered system. We aim to explore whether it is better to deploy a small number of large SIM-aided APs or a large number of small SIM-aided APs when the total number of meta-atoms is limited. It is clear that when the total number of meta-atoms in the system is limited, deploying more APs does not necessarily result in better system performance; there is an optimal number of AP deployments. When the total number of meta-atoms is $N_{\rm{total}} = 300$, the sum rate performance is optimal when deploying 6 APs. Interestingly, if only SIM phase shift optimization is performed without power control, fewer APs are always beneficial. Moreover, as the number of APs increases, the performance gap between using the AGA scheme and the NUA scheme increases. For example, when $L=6$ and $L=25$, the performance of the AGA compared to the NUA improves by 19\% and 200\%, respectively. Specifically, when $L=25$ and the number of meta-atoms per layer of SIM is $N=6$, optimizing only the SIM phase shift results in worse performance than optimizing only power control. This is because when the number of meta-atoms per layer of SIM is too small, it cannot provide sufficient array gain and beamforming capability. This reveals that in SIM-aided CF mMIMO systems, when the number of SIM meta-atoms is limited, we need to reasonably design the number of APs to achieve the maximum system performance improvement. Additionally, with an increasing number of APs, the importance of AP antenna and UE association becomes more evident.

\begin{figure}[t]
\centering
\includegraphics[scale=0.55]{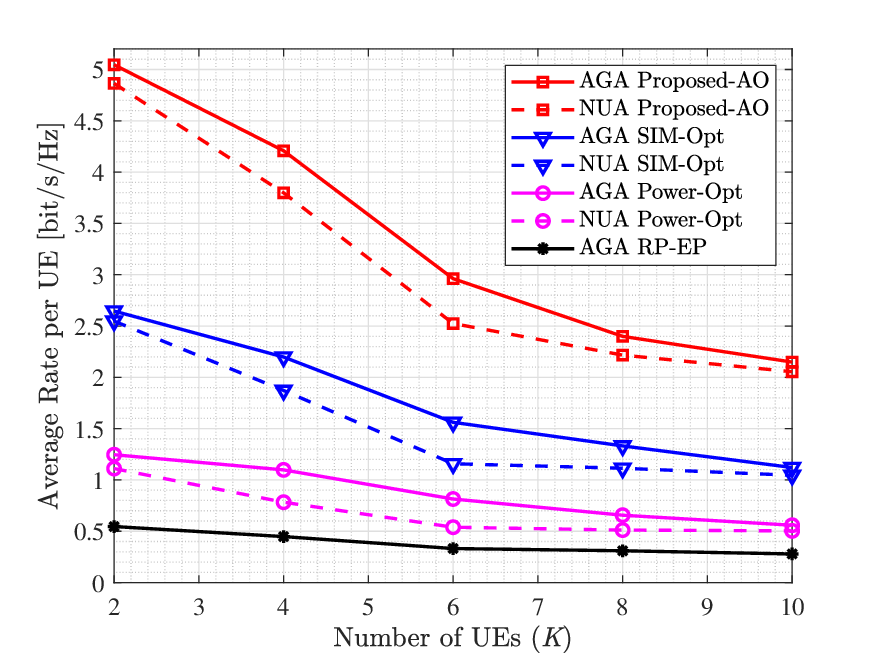}\vspace{-0.2cm}
\caption{Sum rate against different numbers of UEs ($L = 6$, $U = 2$, $M = 2$, $N = 25$, ${d_x} = {d_y} = {\lambda  \mathord{\left/
 {\vphantom {\lambda  2}} \right.
 \kern-\nulldelimiterspace} 2}$).}\label{Fig_K}\vspace{-0.2cm}
\end{figure}

Fig.~\ref{Fig_K} illustrates the average rate per UE against different numbers of UEs with different design schemes of the considered system. It is clear that as the number of UEs increases, the average rate for each UE decreases, but the sum rate of the system increases thanks to the multiuser multiplexing gain. Furthermore, within a reasonable range of UE numbers, the performance of our proposed scheme consistently surpasses that of various scenarios without AP antenna and UE association, demonstrating the practicality and scalability of our proposed algorithm framework. Additionally, as the number of UEs increases, the gap between the proposed AGA and NUA first widens and then narrows. For example, when $L=2$, $L=6$, and $L=10$, the performance gaps between the AGA proposed AO-based framework and NUA proposed AO-based framework are 4\%, 17\%, and 5\%, respectively. This indicates that the difference between the number of UEs and the number of antennas per AP needs to be within a reasonable range; if it is too large or too small, the benefits of AP antenna and UE association are diminished.

\begin{figure}[t]
\centering
\includegraphics[scale=0.55]{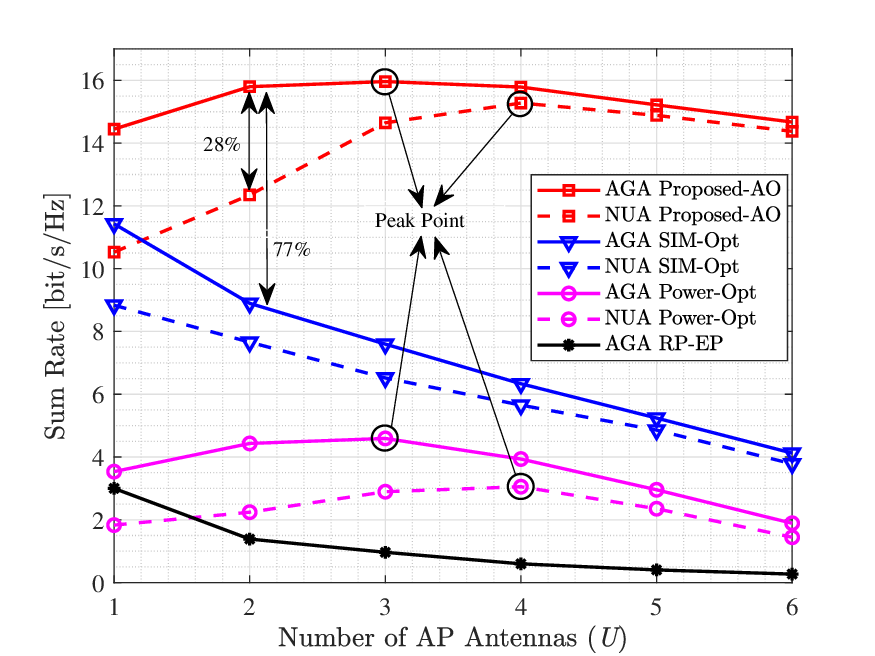}\vspace{-0.2cm}
\caption{Sum rate against different numbers of AP antennas ($L = 6$, $M = 2$, $N = 25$, $K = 4$, ${d_x} = {d_y} = {\lambda  \mathord{\left/
 {\vphantom {\lambda  2}} \right.
 \kern-\nulldelimiterspace} 2}$).}\label{Fig_U}\vspace{-0.2cm}
\end{figure}

\begin{figure}[t]
\centering
\includegraphics[scale=0.55]{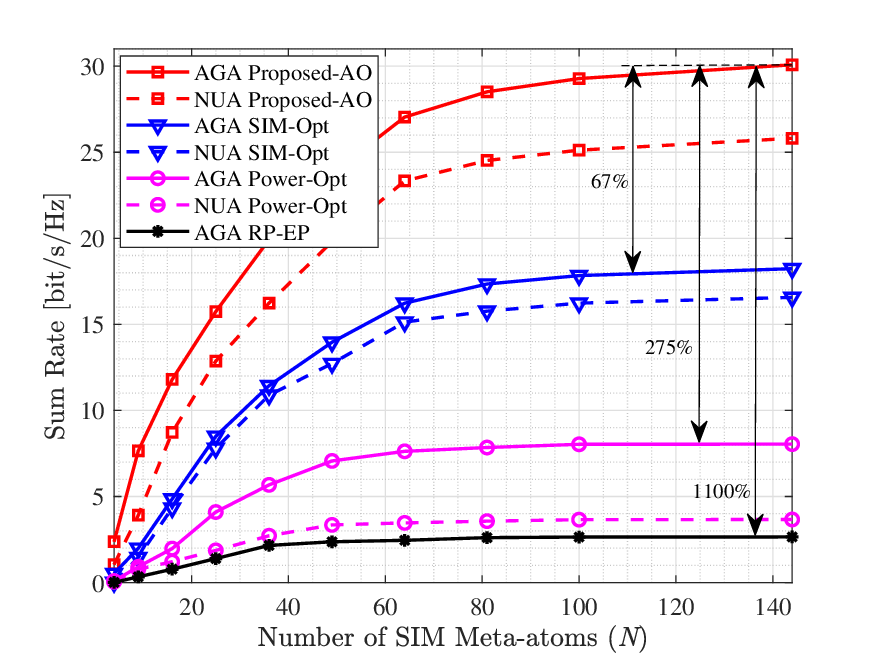}\vspace{-0.2cm}
\caption{Sum rate against different numbers of SIM meta-atoms per layer ($L = 6$, $U = 2$, $M = 2$, $K = 4$, ${d_x} = {d_y} = {\lambda  \mathord{\left/
 {\vphantom {\lambda  2}} \right.
 \kern-\nulldelimiterspace} 2}$).}\label{Fig_N}\vspace{-0.2cm}
\end{figure}

\begin{figure}[t]
\centering
\includegraphics[scale=0.55]{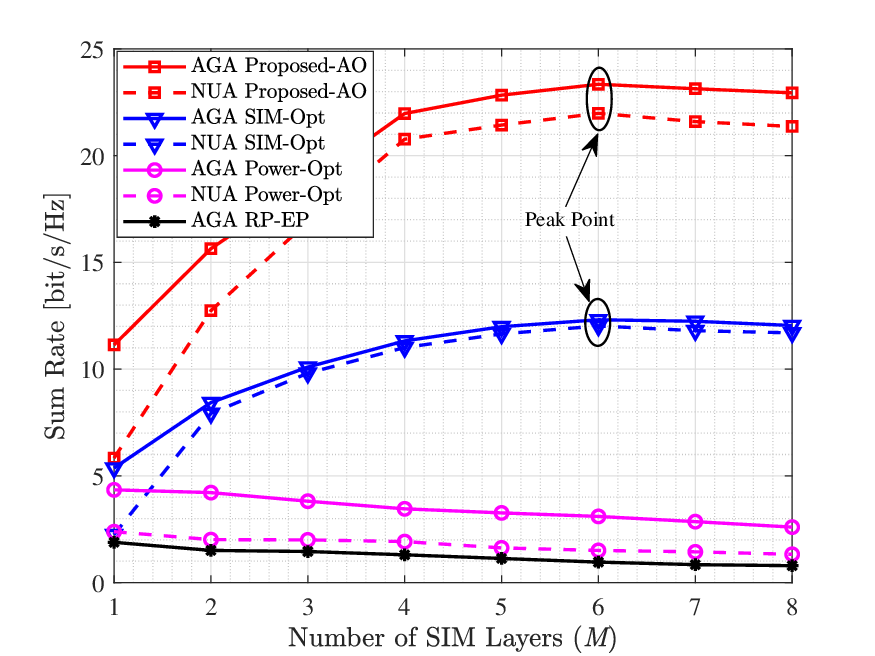}\vspace{-0.2cm}
\caption{Sum rate against different numbers of SIM layers ($L = 6$, $U = 2$, $N = 25$, $K = 4$, ${d_x} = {d_y} = {\lambda  \mathord{\left/
 {\vphantom {\lambda  2}} \right.
 \kern-\nulldelimiterspace} 2}$).}\label{Fig_M}\vspace{-0.2cm}
\end{figure}

Fig.~\ref{Fig_U} illustrates the sum rate against different numbers of AP antennas with different design schemes of the considered system. When AP antenna $U=2$, our proposed AGA AO-based framework improves system performance by 28\%, 77\%, and 256\% compared to the NUA AO-based framework, optimizing only the SIM phase shifts and only the AP antenna power, respectively. It is interesting that as the number of AP antennas increases, the performance of the SIM-aided CF mMIMO system with the proposed AO-based framework first improves and then declines, exhibiting a peak point. Specifically, if only SIM phase shift optimization is adopted, increasing the number of AP antennas results in decreased system performance, which is different from traditional MIMO systems, where increasing the number of antennas continuously enhances system performance. 
This is because, with the introduction of SIMs that have a large number of meta-atoms, especially when AP antennas only control the power of a single UE data stream with equal power, increasing the number of antennas does not enhance spatial diversity gain, array gain, or beamforming capability. 
On the contrary, this leads to a decrease in power per antenna while requiring the AP to transmit more data streams, thereby increasing inter-user interference and optimization complexity.
It reveals that in SIM-aided CF mMIMO systems, it is necessary to design the number of AP antennas reasonably to achieve optimal system performance while avoiding increased complexity and performance degradation.

\subsection{Impact of SIM Parameters}
Fig.~\ref{Fig_N} illustrates the sum rate against different numbers of SIM meta-atoms per layer with different design schemes of the considered system. It is clear that as the number of meta-atoms per layer increases, system performance improves; however, when the number of meta-atoms is extremely large, further increases in the number of meta-atoms result in diminishing performance gains. After performance stabilizes, the proposed AGA-based AO scheme shows performance improvements of 67\%, 275\%, and 1100\% compared to optimizing SIM phase shifts only, optimizing power only, and using random phases with equal power, respectively. Additionally, we observe that without SIM phase shifts optimization, when the number of SIM meta-atoms $N = 64$, the system sum rate nearly ceases to increase. However, adopting SIM phase shifts optimization, the sum rate begins to stabilize when the number of SIM meta-atoms $N = 144$. This indicates that appropriate SIM phase shift optimization is necessary to fully harness the potential of a large number of SIM meta-atoms and achieve further improvements in the performance of the SIM-aided CF mMIMO system.

\begin{figure}[t]
\centering
\includegraphics[scale=0.55]{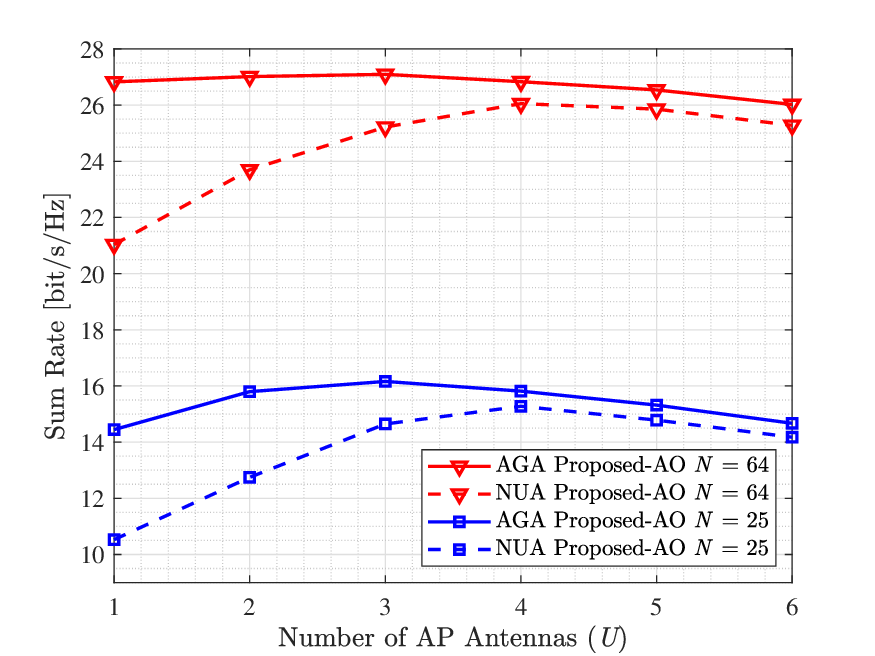}\vspace{-0.2cm}
\caption{Sum rate against different numbers of AP antennas and SIM meta-atoms per layer ($L = 6$, $M = 2$ $K = 4$, ${d_x} = {d_y} = {\lambda  \mathord{\left/
 {\vphantom {\lambda  2}} \right.
 \kern-\nulldelimiterspace} 2}$).}\label{Fig_U_N}\vspace{-0.2cm}
\end{figure}

Fig.~\ref{Fig_M} illustrates the sum rate against different numbers of SIM metasurface layers with different design schemes of the considered system. It is clear that as the number of SIM metasurface layers increases, the system sum rate initially increases. However, after the number of metasurface layers reaches $M = 6$, further increasing the number of layers no longer improves performance; instead, it starts to decline. This indicates that there is an optimal number of SIM layers that maximizes the system sum rate. This phenomenon occurs because there is a transmission penetration loss coefficient $\mathbf{W}$ between two layers of SIM. When the number of layers is relatively small, the performance gain brought by the additional meta-atoms per layer outweighs the degradation caused by the penetration loss between the layers, resulting in performance improvement. However, when the number of layers exceeds $M = 6$, the performance gain from adding more layers cannot offset the penetration loss between the layers, leading to a decline in performance. Furthermore, if the SIM phase shifts are not optimized and only antenna power control is adopted, the system performance continues to decline as the number of SIM layers increases. This is because random phase shifts cannot fully utilize the beamforming capabilities of the large number of SIM meta-atoms. Additionally, the results show that as the number of layers increases, the performance gap between the AGA algorithm and the NUA algorithm for AP antennas decreases. This reveals that increasing the number of SIM layers can enhance the joint beamforming capability of SIM for multi-user signals.

Fig.~\ref{Fig_U_N} illustrates the sum rate against different numbers of AP antennas and SIM meta-atoms per layer of the considered system. It is evident that when the number of SIM meta-atoms is $N = 25$, increasing the number of AP antennas initially enhances system performance, which then declines, showing a peak. However, when the number of SIM meta-atoms is large, e.g. $N = 64$, increasing the number of AP antennas results in only a slight performance improvement, and when the number of antennas exceeds three, it leads to a decline in system performance. It indicates that when there are sufficient SIM meta-atoms, a large number of AP antennas is no longer needed to improve system performance. Instead, increasing the number of antennas leads to smaller power allocation per antenna and increased complexity, making it easier to fall into local optima. Additionally, when the number of antennas equals the number of UEs (both four), the performance gap between the AGA scheme and the NUA scheme becomes very small. This suggests that in practice when the number of AP antennas is less than the number of UEs, the AP antenna and UE association scheme becomes more crucial.

\begin{figure*}[t!]
\normalsize
\setcounter{mytempeqncnt}{1}
\setcounter{equation}{26}
\begin{align}\label{Them2}
\frac{{\partial {\gamma _k}}}{{\partial \varphi _{l,m}^n}} = \frac{1}{{{\xi _k}}} \cdot \frac{{\partial {{\left| {\sum\limits_{l = 1}^L {{\bf{h}}_{l,k}^H{{\bf{G}}_l}{{\bf{W}}_{l,1}}{{\bf{P}}_l}{{\bf{a}}_{l,k}}} } \right|}^2}}}{{\partial \varphi _{l,m}^n}} - \frac{{{{\left| {\sum\limits_{l = 1}^L {{\bf{h}}_{l,k}^H{{\bf{G}}_l}{{\bf{W}}_{l,1}}{{\bf{P}}_l}{{\bf{a}}_{l,k}}} } \right|}^2}}}{{\xi _k^2}} \cdot \mathop \sum \limits_{j = 1,j \ne k}^K \frac{{\partial {{\left| {\sum\limits_{l = 1}^L {{\bf{h}}_{l,k}^H{{\bf{G}}_l}{{\bf{W}}_{l,1}}{{\bf{P}}_l}{{\bf{a}}_{l,j}}} } \right|}^2}}}{{\partial \varphi _{l,m}^n}},
\end{align}
\setcounter{equation}{27}
\hrulefill
\end{figure*}

\section{Conclusion}
In this paper, we introduce a novel SIM-aided CF mMIMO paradigm, which reduces the hardware design complexity and computational load of APs. This paradigm eliminates the need for baseband digital encoding, enabling precoding at the speed of light in the natural EM regime. 
Considering the characteristics of SIM, we first formulate the problem of joint AP antenna-UE association and precoding at APs and SIMs to maximize the system sum rate. Then, we propose an efficient two-stage signal processing framework to solve the non-convexity and high complexity problem. In the first stage, we propose an AGA algorithm to manage inter-user interference. In the second stage, we propose an AO-based algorithm that decomposes the joint power and wave-based precoding optimization into two distinct sub-problems. The complex quadratic transform method is utilized for AP antenna power control, while the PGA algorithm is applied to obtain suboptimal solutions for SIM wave-based precoding. 

The results indicate that the proposed two-stage signal processing framework can efficiently improve the sum rate of the SIM-aided CF mMIMO system, which surpasses various existing benchmark schemes. Specifically, we find that when the total number of SIM meta-atoms is limited, deploying more low-end APs is not necessarily better; there is an optimal number of deployments. In our scenario, with a total of 300 meta-atoms, the optimal number of APs is 6. Also, the results show that a 6-layer SIM with 25 meta-atoms per layer having half-wavelength element spacing achieves the maximum sum rate. Besides, when the number of meta-atoms configured in the SIM is sufficient to achieve a stable sum rate, the proposed scheme shows a 275\% performance improvement compared to adopting only the antenna power control scheme.
Finally, deploying SIMs with a substantial number of meta-atoms can serve as an effective alternative to APs and AP antennas, thereby enhancing the performance of CF mMIMO systems.

\begin{figure*}[t!]
\normalsize
\setcounter{mytempeqncnt}{1}
\setcounter{equation}{29}
\begin{align}\label{Them4}
\frac{{\partial {{\left| {\sum\limits_{l = 1}^L {{\bf{h}}_{l,k}^H{{\bf{G}}_l}{{\bf{W}}_{l,1}}{{\bf{P}}_l}{{\bf{a}}_{l,k}}} } \right|}^2}}}{{\partial \varphi _{l,m}^n}} &= \frac{{\partial {{\left| {\sum\limits_{n = 1}^N {{e^{j\varphi _{l,m}^n}}} \sum\limits_{l = 1}^L {{\bf{h}}_{l,k}^H{\bf{b}}_{l,m}^n{{\left( {{\bf{q}}_{l,m}^n} \right)}^{\rm{H}}}{{\bf{W}}_{l,1}}{{\bf{P}}_l}{{\bf{a}}_{l,k}}} } \right|}^2}}}{{\partial \varphi _{l,m}^n}} \notag\\
&= 2 \cdot \frac{{\partial \Re \left[ {\left( {{e^{j\varphi _{l,m}^n}}\sum\limits_{l = 1}^L {{\bf{h}}_{l,k}^H{\bf{b}}_{l,m}^n{{\left( {{\bf{q}}_{l,m}^n} \right)}^{\rm{H}}}{{\bf{W}}_{l,1}}{{\bf{P}}_l}{{\bf{a}}_{l,k}}} } \right){{\left( {\sum\limits_{l = 1}^L {{\bf{h}}_{l,k}^H{{\bf{G}}_l}{{\bf{W}}_{l,1}}{{\bf{P}}_l}{{\bf{a}}_{l,k}}} } \right)}^{\rm{H}}}} \right]}}{{\partial \varphi _{l,m}^n}} \notag\\
&= 2 \cdot \Im \left[ {{{\left( {{e^{j\varphi _{l,m}^n}}\sum\limits_{l = 1}^L {{\bf{h}}_{l,k}^H{\bf{b}}_{l,m}^n{{\left( {{\bf{q}}_{l,m}^n} \right)}^{\rm{H}}}{{\bf{W}}_{l,1}}{{\bf{P}}_l}{{\bf{a}}_{l,k}}} } \right)}^{\rm{H}}}\left( {\sum\limits_{l = 1}^L {{\bf{h}}_{l,k}^H{{\bf{G}}_l}{{\bf{W}}_{l,1}}{{\bf{P}}_l}{{\bf{a}}_{l,k}}} } \right)} \right].
\end{align}
\setcounter{equation}{30}
\hrulefill
\end{figure*}

\begin{appendices}
\section{Proof of Theorem 1}
This appendix calculates the formula \eqref{deta_R} in Theorem \ref{them}. To start with, we note that the gradient of $R_{\rm{sum}}$ with respect to $\varphi^n_{l,m}$ can be written as 
\begin{align}\label{Them1}
\setcounter{equation}{25}
    \frac{{\partial {R_{{\rm{sum}}}}}}{{\partial \varphi _{l,m}^n}} \!=\! {\log _2}e \!\cdot\!\! \sum\limits_{k = 1}^K \!{\frac{1}{{1 \!+\! {\gamma _k}}} \cdot } \frac{{\partial {\gamma _k}}}{{\partial \varphi _{l,m}^n}}, \forall l \!\in\! {\cal L},\forall m \!\in\! {\cal M},\forall n \!\in\! {\cal N}.
\end{align}
Based on the standard quotient rule for derivatives, we derive $\frac{{\partial {\gamma _k}}}{{\partial \varphi _{l,m}^n}}$ in \eqref{Them1} as \eqref{Them2} at the top of this page, where ${\xi _k}$ is given by 

\begin{align}\label{xi_k}
\setcounter{equation}{27}
    \xi_k  = \mathop \sum \limits_{j = 1,j \ne k}^K {\left| {\sum\limits_{l = 1}^L {{\bf{h}}_{l,k}^H{{\bf{G}}_l}{{\bf{W}}_{l,1}}{{\bf{P}}_l}{{\bf{a}}_{l,j}}} } \right|^2} + \sigma _k^2.
\end{align}

With the help of \eqref{Them2}, the partial derivative $\frac{{\partial {R_{{\rm{sum}}}}}}{{\partial \varphi _{l,m}^n}}$ in \eqref{Them1} can be further derived and simplified to
\begin{align}\label{Them3}
\frac{{\partial {R_{{\rm{sum}}}}}}{{\partial \varphi _{l,m}^n}} &= {\log _2}e \cdot \sum\limits_{k = 1}^K {{\delta _k}}  \cdot \left( {\frac{{\partial {{\left| {\sum\limits_{l = 1}^L {{\bf{h}}_{l,k}^H{{\bf{G}}_l}{{\bf{W}}_{l,1}}{{\bf{P}}_l}{{\bf{a}}_{l,k}}} } \right|}^2}}}{{\partial \varphi _{l,m}^n}}} \right.\notag\\
&\left. { - {\gamma _k}\mathop \sum \limits_{j = 1,j \ne k}^K \frac{{\partial {{\left| {\sum\limits_{l = 1}^L {{\bf{h}}_{l,k}^H{{\bf{G}}_l}{{\bf{W}}_{l,1}}{{\bf{P}}_l}{{\bf{a}}_{l,j}}} } \right|}^2}}}{{\partial \varphi _{l,m}^n}}} \right),
\end{align}
where ${\delta _k} = \frac{1}{{{\xi _k}}}$ which is defined in \eqref{dalta}.

Next, the primary challenge of \eqref{Them3} is determining the partial derivative of ${{{| {\sum\limits_{l = 1}^L {{\bf{h}}_{l,k}^H{{\bf{G}}_l}{{\bf{W}}_{l,1}}{{\bf{P}}_l}{{\bf{a}}_{l,k}}} } |^2}}}$ with respect to ${e^{j\varphi _{l,m}^n}}$. It is important to note that ${\sum\limits_{l = 1}^L {{\bf{h}}_{l,k}^H{{\bf{G}}_l}{{\bf{W}}_{l,1}}{{\bf{P}}_l}{{\bf{a}}_{l,k}}} }$ is linear with respect to ${e^{j\varphi _{l,m}^n}}$. Therefore, for each given pair of $k,j \in \mathcal{K}$, we can get \eqref{Them4} at the top of the next page, where ${{\bf{b}}_{l,m}^n}$ and ${{{( {{\bf{q}}_{l,m}^n} )}^{\rm{H}}}}$ are defined in \eqref{B} and \eqref{Q}, respectively.
Finally, the proof is completed by substituting \eqref{Them4} into \eqref{Them3}.
\end{appendices}

\bibliographystyle{IEEEtran}
\bibliography{IEEEabrv,Ref}

\end{document}